\documentclass[10pt, oneside]{article}
\usepackage[left=35mm, right=35mm, top=35mm, bottom=40mm]{geometry}
\usepackage{tikz}
\usepackage{color, graphicx}
\usepackage{amsmath, amssymb, amsfonts, mathtools}
\usepackage[square, authoryear]{natbib}
\usepackage{enumitem}
\usepackage{authblk}
\usepackage{titlesec}
\usepackage[labelsep=period,labelfont=bf,font=small]{caption}
\usepackage{bm}
\usepackage{float}
\usepackage[hidelinks]{hyperref}
\usepackage{afterpage}
\usepackage{xpatch}
\usepackage{subfigure}
\usepackage[section]{placeins}
\xpatchcmd{\author}{\relax#1\relax}{\relax\detokenize{#1}\relax}{}{}

\graphicspath{{Figures/}}

\author[2,3]{Shengnan Lyu}
\author[3]{Christian W\"ulker}
\author[3]{Yuqing Pan}
\author[1]{Amitesh S. Jayaraman}
\author[4]{Jianhao Zheng}
\author[4]{Yilin Cai}
\author[1,3]{Gregory S.\ Chirikjian\thanks{Corresponding author. E-mail: \texttt{mpegre@nus.edu.sg}}}

\affil[1]{\small{Department of Mechanical Engineering, National University of Singapore, Singapore}}
\affil[2]{\small{School of Mechanical Engineering and Automation, Beihang University, Beijing, China}}
\affil[3]{\small{Department of Mechanical Engineering, Johns Hopkins University, Baltimore, MD, USA}}
\affil[4]{\small{School of Mechanical Engineering, Shanghai Jiao Tong University, Shanghai, China}}

\setlist[enumerate]{wide=\parindent}

\makeatletter
\def\moverlay{\mathpalette\mov@rlay}
\def\mov@rlay#1#2{\leavevmode\vtop{%
   \baselineskip\z@skip \lineskiplimit-\maxdimen
   \ialign{\hfil$\m@th#1##$\hfil\cr#2\crcr}}}
\newcommand{\charfusion}[3][\mathord]{
    #1{\ifx#1\mathop\vphantom{#2}\fi
        \mathpalette\mov@rlay{#2\cr#3}
      }
    \ifx#1\mathop\expandafter\displaylimits\fi}
\makeatother

\newcommand{\IR}{\mathbb{R}}

\title{\Large\textbf{Cross-Modal Fusion Between Data in SAXS and Cryo-EM for Biomolecular Structure Determination}}
\date{\small\today}
\begin{document}

\maketitle

\begin{abstract}
Cryo-Electron Microscopy (cryo-EM) has become an extremely powerful method for resolving structural details of large biomolecular complexes. However, challenging problems in single-particle methods remain open because of (1) the low signal-to-noise ratio in EM; and (2) the potential anisotropy and lack of coverage of projection directions relative to the body-fixed coordinate system for some complexes. Whereas (1) is usually addressed by class averaging (and increasingly due to rapid advances in microscope and sensor technology), (2) is an artifact of the mechanics of interaction of biomolecular complexes and the vitrification process. In the absence of tilt series, (2) remains a problem, which is addressed here by supplementing EM data with Small-Angle X-Ray Scattering (SAXS). Whereas SAXS is of relatively low resolution and contains much lower information content than EM, we show that it is nevertheless possible to use SAXS to fill in blind spots in EM in difficult cases where the range of projection directions is limited.

\end{abstract}

\section{Introduction}
\label{sec:introduction}
The goal of this paper is the fusion of information from \emph{cryo-electron microscopy} (cryo EM) and \emph{small-angle X-ray scattering} (SAXS) in order to exploit the synergies of both techniques. Both of these methods are experimental biophysical techniques used to glean information about the structure of (large) biomolecular complexes in the native state. The data acquisition in these methods can be briefly summarized as follows: in so-called single-particle EM, we are presented with noisy projections of the electron density of a molecular structure from a finite number of \textit{a priori} unknown different directions relative to its body-fixed frame. The three-dimensional shape is to be reconstructed from these projection data, \textit{i.\,e.}, EM is a tomographic imaging technique. SAXS data acquisition, on the other hand, can be interpreted as taking the electron density of the structure\,---\,or rather the result of a convolution of the density with a reflected version of itself\,---\,and subsequent averaging over concentric spheres centered at the origin. This is equivalent to square averaging the Fourier spectrum of the electron density on these spheres. For a detailed discussion of the mathematical aspects of SAXS and EM data acquisition, we refer the reader to \cite{dong_kim_chirikjian, afsari_kim_chirikjian}.

The information content and resolution of EM projections, either after class averaging or due to new advances in imaging technology, are both significantly higher than SAXS. Nevertheless, we make the case that there are potential benefits to fuse these information sources.

The main tool that we use in this work to fuse the information from EM and SAXS is the well-known \emph{Fourier slice theorem}. In two dimensions, this theorem states that the results of the following two calculations are equal:

\begin{enumerate}[wide, labelwidth=!, labelindent=0pt]
 \item Taking a two-dimensional function $f(x,y)$, projecting it onto some (one-dimensional) line, and doing a Fourier transform of that projection along that line.
 \item Taking the function $f(x,y)$, performing a two-dimensional Fourier transform, and then slicing it through its origin with a line parallel to the projection line mentioned above.
\end{enumerate}

This theorem directly generalizes into higher dimensions; particularly, it also holds in the three-dimensional case.

The starting point of this work is the Fourier slice theorem: we can interpret EM data as spectral data given on a finite number of planes passing through the origin as parts of the Fourier spectrum of the electron density of the molecular structure of interest. The main issue here is that there might be significantly large gaps between these planes in the spectral EM data, because the number of directions in which projection data acquired in EM can be very limited. This makes reconstruction via the two-dimensional inverse Fourier transform potentially problematic. As indicated above, SAXS adds to the spectral EM data, for here the data can be seen as constant values on concentric spheres centered at the origin of the Fourier space, resulting from square averaging of the spectrum over these spheres. Therefore, both EM and SAXS data can be considered in this context as partial information on the Fourier spectrum of the electron density to be reconstructed.

The remainder of this paper is organized as follows: in Section \ref{sec:theory}, we derive our technique for SAXS-EM data fusion and in Section \ref{Roots} we explore some properties of the multiple admissible solutions for the reconstructed data of the SAXS-EM fusion procedure. Section \ref{sec:numerical_experiments} presents numerical results, validating the practical applicability of our method. In Section \ref{sec:discussion_and_conclusion}, we discuss our results and give an outlook on future developments.

\section{Literature Review}
Macromolecular structure determination is of central importance in biology because of the close relationship between shape and function. As the fields of biology and biophysics increasingly focus on the study of cellular phenomena, understanding the structure and motion of large biomolecular complexes is becoming critical. Large biomolecular complexes bridge the length scales of single-molecule studies and cellular phenomena.

Two of the main methods used in this context are small-angle X-ray scattering (SAXS)
\cite{ stuhrmann1970interpretation, feigin1987structure, blanchet2013small, svergun2003small,
dong2015computational} and cryo-electron microscopy (cryo EM) \cite{crowther1970reconstruction, van1981use, penczek1992three,  frank2006three,  wang2006cryo, scheres2012relion}. SAXS is known for its relatively low resolution and low information content in comparison to cryo-EM. Nevertheless, the information provided in these methods can be used to complement each other provided that the structural features in both experimental conditions are compatible \cite{kim2017cross}.

The data acquisition processes in these methods are also quite different. A SAXS experiment is fast to perform (given access to a synchrotron) and requires relatively little sample preparation. But it provides very low resolution data. In contrast, a cryo-EM experiment can usually be performed locally on microscopes owned by most research-intensive universities, and provides large amounts of data. But EM needs more effort for sample preparation and moreover, post-processing computations are necessary to classify, denoise and/or class-average images \cite{ bhamre2016denoising, park2014assembly, park2011stochastic, penczek2002three, schatz1990invariant, sigworth1998maximum}. After this, the images are aligned to yield a 3D density map at high resolution \cite{penczek1996common, scheres2005maximum, shkolnisky2012viewing, singer2011viewing, wang2006cryo}.

Usually either SAXS or cryo-EM is used, but here we investigate the ``fusion" of structural information obtained from SAXS and EM experiments. Although the type of data and the data acquisition processes in these methods are quite different, they do share some common features. Both SAXS and EM differ from crystallography in that no crystallization is needed (at the expense of lower resolution). The advantage is that finding appropriate crystallization conditions in many cases is either a very lengthy process or the process may capture the molecules in conditions that are not biologically relevant.

That said, in SAXS experiments, the biomolecular complexes of interest are in solution and are exposed to X-ray beams, which leads to the scattering of X-ray photons. Unlike in X-ray crystallography where the macromolecules are regularly positioned and oriented, in SAXS the molecules can move and rotate randomly in solution. Hence, roughly speaking, the low resolution in SAXS can be attributed to the fact that the SAXS data is the spherical average of the scattering pattern of the complex under study.

In cryo-EM, similar to standard tomography, we obtain numerous two-dimensional (2D) projections of a 3D complex. A specimen grid, with the molecules of interest in it, is prepared and exposed to high-energy electron beams to obtain millions of projected 2D images. However, in contrast to tomography, the projections are at random (unknown) directions; moreover, the projections are extremely noisy. As a result, the 3D volume reconstruction in cryo-EM involves complicated postprocessing and still the resolution is not very high as compared to crystallographic structures (although it is much higher than SAXS).

Both SAXS and EM have been successfully used for the structural study of large biomolecular complexes. However, as explained earlier, they have their own disadvantages for the study of function–structure relationships. To better investigate the relationship between the function and the shape of a biomolecular complex, it may be desirable to fuse the information from these two experimental data modalities, which is the goal of this paper.

\section{Theory}
\label{sec:theory}
In this section, we elaborate on the theory of SAXS-EM data fusion in the two-dimensional and three-dimensional case. We consider the two-dimensional case first, because we will later see that our idea directly generalizes to the three-dimensional case, which simplifies the derivation.

\subsection{The two-dimensional case (For conceptualization)}

Though proteins are 3D structures, before going into the mathematical details of 3D case, we first develop a conceptual framework illustrated with a 2D toy model.

\begin{figure}[!h]
\centering
\includegraphics[width = 7cm]{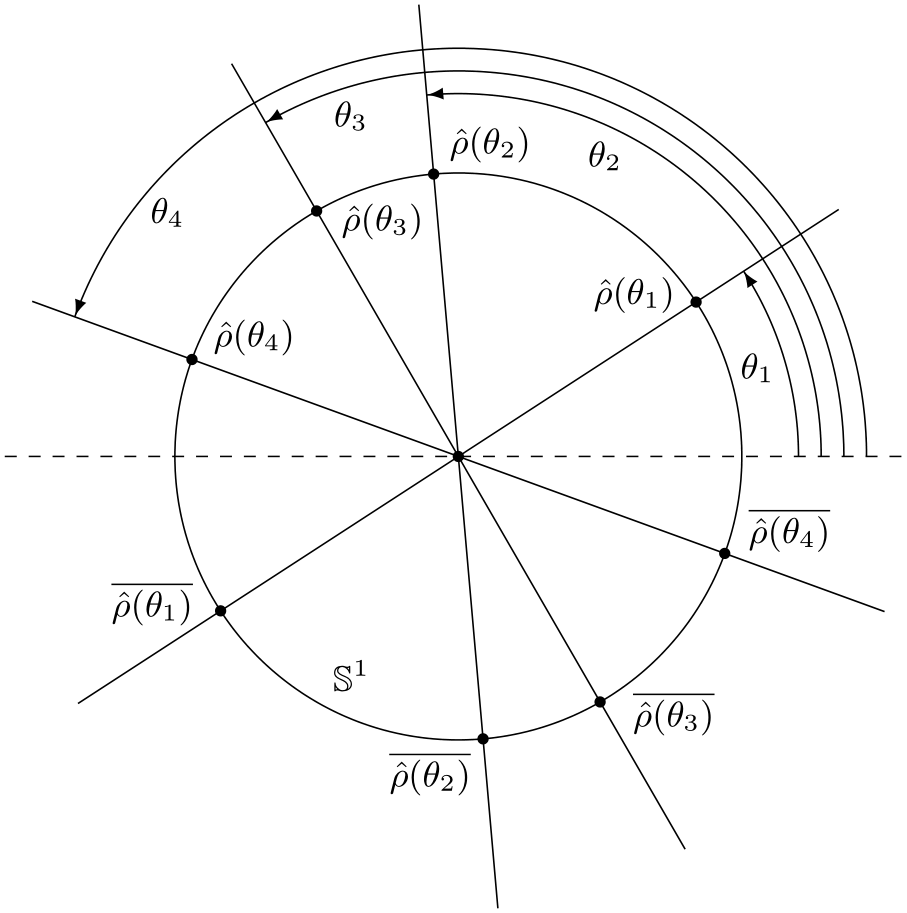}
\caption{SAXS-EM data in the two-dimensional setting (conceptual plot).}
\label{fig:2d}
\end{figure}

In the two-dimensional case, the task is to reconstruct the unknown electron density $\rho(x,y)$ of a fictitious 2D structure of interest from partial Fourier data. The idea of information fusion is to combine the information from both EM and SAXS to fill in the missing values of the Fourier spectrum $\hat{\rho}(\omega_1, \omega_2)$ in the plane, so that $\rho$ can be reconstructed via a two-dimensional inverse Fourier transform. Figure \ref{fig:2d} shows the basic two-dimensional setting. In Fourier space, from EM and the Fourier slice theorem, the values of the spectrum $\hat{\rho}$ on each line passing through the origin are known. On the other hand, the values of the square average of $\hat{\rho}$ on each of the concentric circles are known from SAXS. For simplicity, Fig.\,\ref{fig:2d} shows only one such circle, with radius equal to one. Many such circles will be used to cover the plane in Fourier space.

\FloatBarrier

Switching to polar coordinates, $r$ and $\theta$, we see that the value of $\hat{\rho}(r\cos\theta_i,r\sin\theta_i)$ at the intersection between each line and the circle is known, because from the Fourier slice theorem, the EM data can be seen as the values of $\hat{\rho}$ on lines passing through the origin, perpendicular to the respective projection direction (in 3D the slice would be a plane and the projection direction would be normal to this plane). The sought-after density function $\rho:\mathbb{R}^2 \rightarrow \mathbb{R}_{\geq 0}$ has the Fourier transform

$$\hat{\rho}(\bm{\omega}) \,=\, \int_{\mathbb{R}^2} \rho(\bm{x}) \mathrm{e}^{- \mathrm{i} \bm{\omega}^\textnormal{T} \bm{x}} \mathrm{d} \bm{x}.$$

The original function $\rho$ can be reconstructed from its Fourier transform via the inverse transform,

\begin{equation}\label{eq:2d_Fourier_inverse}
\rho(\bm{x}) \,=\, (2\pi)^{-2} \int_{\mathbb{R}^2} \hat{\rho}(\bm{\omega}) \mathrm{e}^{\mathrm{i} \bm{\omega}^\textnormal{T} \bm{x}} \mathrm{d} \bm{\omega}.
\end{equation}

In polar coordinates, the frequency vector can be written as $\bm{\omega} = r \cdot (\cos\theta, \sin \theta)^\mathrm{T}$, where $r = \|\bm{\omega}\|_2$ is the Euclidean 2-norm of $\bm{\omega}$. The plane is divided into an infinite number of circles with radius $r \in \mathbb{R}_{\geq 0}$. Without loss of generality, let us choose in Fig.\,\ref{fig:2d} the unit circle $\mathbb{S}^1$ with radius $r = 1$, so that $\bm{\omega} = (\cos\theta, \sin\theta)^\textnormal{T}$. In this case, we can use the short hand $\hat{\rho}(\bm{\omega}) \equiv \hat{\rho}(\theta)$, so that $\hat{\rho}(\theta)$ describes a function on the unit circle in the $\bm{\omega}$-plane. Since $-(\cos\theta,\sin\theta) = (\cos(\theta+\pi),\, \sin(\theta+\pi))$, we get directly from the definition of the Fourier transform

\begin{equation}\label{eq:plus_pi}
\overline{\hat{\rho}(\theta)} \,=\, \hat{\rho}(\theta+\pi).
\end{equation}

From the various projections in EM, only the values of the function $\hat{\rho}$ along specific lines in $\mathbb{R}^2$ is obtained, and these lines are not uniformly spaced with respect to their angle. Our goal is to interpolate the data so as to match the projection data of the experiment, while also getting meaningful ``in-between'' values of the function $\hat{\rho}$ on the whole circle. The SAXS experiment provides the square average of the function $\hat{\rho}$ along the circle, which is the information that we shall incorporate in the interpolation process.

For the unit circle $\mathbb{S}^1$ with radius $r = 1$, the 2D version of the SAXS experiment gives the single value

\begin{equation}
I \,\coloneqq\, \int_{\mathbb{S}^1} |\hat{\rho}(\theta)|^2 \mathrm{d} \theta \,=\, \int_{\mathbb{S}^1} \hat{\rho}(\theta) \overline{\hat{\rho}(\theta)} \mathrm{d}\theta.
\label{eq_2}
\end{equation}

On the other hand, data $\hat{\rho}(\theta_i)$ and $\hat{\rho}(\pi + \theta_i) = \overline{\hat{\rho}(\theta_i)}$ at specific points $\theta_i$, $i = 1,2, \dots, m$, on the unit circle are provided by EM, where $m$ is the number of different projection angles. We approximate the function $\hat{\rho}$ with a truncated Fourier series (\textit{i.\,e.}, a trigonometric polynomial),

\begin{equation}
\hat{\rho}(\theta) \,=\, \sum_{k=-n}^n \tilde{\rho}_k \mathrm{e}^{\mathrm{i} k \theta},
\label{eq_3}
\end{equation}

\noindent where
\begin{equation}\label{eq:tilde_rho}
\tilde{\rho}_k \,=\, (2\pi)^{-1} \int_0^{2\pi} \hat{\rho}(\theta) \mathrm{e}^{-\mathrm{i}k\theta} \mathrm{d}\theta
\end{equation}
are the (classical) Fourier coefficients of the function $\hat{\rho}$ on the unit circle $\mathbb{S}^1$, and $n \in \{0, 1, 2, \dots\}$ is the approximation order. We note that

\begin{equation}\label{eq:minus_k}
\tilde{\rho}_{-k} \,=\, (-1)^k \, \overline{\tilde{\rho}_k},
\end{equation}

\noindent due to \eqref{eq:plus_pi}.

Now we substitute \eqref{eq_3} into \eqref{eq_2}, and derive

\begin{align}
\label{eq_5}
I
\,&=\, \int_{0}^{2\pi} \bigg(\sum_{k=-n}^{n} \tilde{\rho}_k \mathrm{e}^{\mathrm{i}k\theta}\bigg) \overline{\bigg(\sum_{k'=-n}^n \tilde{\rho}_{k'} \mathrm{e}^{\mathrm{i}k'\theta}\bigg)} \mathrm{d}\theta \\\notag
\,&=\, \sum_{k,k'=-n}^{n} \tilde{\rho}_k \, \overline{\tilde{\rho}_{k'}} \int_{0}^{2\pi} \mathrm{e}^{\mathrm{i}k\theta} \, \mathrm{e}^{-\mathrm{i}k'\theta} \mathrm{d}\theta \\\notag
\,&=\, 2\pi \sum_{k,k'=-n}^{n} \tilde{\rho}_k \, \overline{\tilde{\rho}_{k'}} \, \delta_{k,k'} \\\notag
\,&=\, 2\pi \sum_{k=-n}^{n} |\tilde{\rho}_k|^2 \notag,
\end{align}

\noindent where $\delta_{k,k'}$ is the Kronecker delta symbol, which is equal to one if $k = k'$ and equal to zero otherwise. Equation \eqref{eq_5} is an approximation to the well-known Parseval equality. Now let us bring this equation into matrix-vector notation. To this end, we introduce the vectors

\begin{equation*}
\bm{\tilde{\rho}} \,\coloneqq\,
\begin{pmatrix}
\tilde{\rho}_{-n}, \dots, \tilde{\rho}_{-1} \,|\, \tilde{\rho}_0 \,|\, \tilde{\rho}_{1}, \dots, \tilde{\rho}_n
\end{pmatrix}^\textnormal{T}.
\end{equation*}

$$\bm{\hat{\rho}} =
\begin{pmatrix}
\hat{\rho}(\theta_1), \hat{\rho}(\theta_2), \dots, \hat{\rho}(\theta_{2m})
\end{pmatrix}^\textnormal{T}.$$

With this vector, \eqref{eq_5} attains the form

\begin{equation*}
I \,=\, 2 \pi \bm{\tilde{\rho}}^\textnormal{H} \bm{\tilde{\rho}},
\end{equation*}

\noindent where $\bm{\tilde{\rho}}^\textnormal{H} = \overline{\bm{\tilde{\rho}}^\textnormal{T}}$ is the Hermitian transpose of $\bm{\tilde{\rho}}$.

Now turning back to (\ref{eq_3}), we find that
\begin{equation}
\bm{\hat{\rho}} = A \bm{\tilde{\rho}}.
\label{eq:eq_matrix}
\end{equation}

\noindent where $$A = \begin{pmatrix}
\begin{array}{ccc|c|ccc}
\mathrm{e}^{-\mathrm{i} n \theta_1} & \ldots & \mathrm{e}^{-\mathrm{i} \theta_1} & 1 & \mathrm{e}^{\mathrm{i} \theta_1} & \ldots & \mathrm{e}^{\mathrm{i} n \theta_1} \\
\mathrm{e}^{-\mathrm{i} n \theta_2} & \ldots & \mathrm{e}^{-\mathrm{i} \theta_2} & 1 & \mathrm{e}^{\mathrm{i} \theta_2} & \ldots & \mathrm{e}^{\mathrm{i} n \theta_2} \\
\vdots & & \vdots & \vdots & \vdots & & \vdots \\
\mathrm{e}^{-\mathrm{i} n \theta_{2m}} & \ldots & \mathrm{e}^{-\mathrm{i} \theta_{2m}} & 1 &  \mathrm{e}^{\mathrm{i} \theta_{2m}} & \ldots & \mathrm{e}^{\mathrm{i} n \theta_{2m}}
\end{array}
\end{pmatrix} \in \, \mathbb{C}^{2m \times (2n+1)}.$$

It can be seen from \eqref{eq:tilde_rho} that the unknown $\tilde{\rho}_0$ is real. We bring this unknown quantity to the left-hand side of \eqref{eq:eq_matrix},

\begin{equation}
\bm{\hat{\rho}} - \tilde{\rho}_0\bm{1} = F \bm{\tilde{\rho}'}
\label{eq:system}
\end{equation}

\noindent where $\bm{1}$ is the vector of length $2m$ containing only ones,
\begin{equation*}
F = \begin{pmatrix}
\begin{array}{ccc|ccc}
\mathrm{e}^{-\mathrm{i} n \theta_1} & \ldots & \mathrm{e}^{-\mathrm{i} \theta_1} & \mathrm{e}^{\mathrm{i} \theta_1} & \ldots & \mathrm{e}^{\mathrm{i} n \theta_1} \\
\mathrm{e}^{-\mathrm{i} n \theta_2} & \ldots & \mathrm{e}^{-\mathrm{i} \theta_2} & \mathrm{e}^{\mathrm{i} \theta_2} & \ldots & \mathrm{e}^{\mathrm{i} n \theta_2} \\
\vdots & & \vdots & \vdots & & \vdots \\
\mathrm{e}^{-\mathrm{i} n \theta_{2m}} & \ldots & \mathrm{e}^{-\mathrm{i} \theta_{2m}} & \mathrm{e}^{\mathrm{i} \theta_{2m}} & \ldots & \mathrm{e}^{\mathrm{i} n \theta_{2m}}
\end{array}
\end{pmatrix} \in\ \mathbb{C}^{2m \times 2n},
\end{equation*}
\begin{equation}
\bm{\tilde{\rho}'} = \begin{pmatrix}
\tilde{\rho}_{-n} & \cdots & \tilde{\rho}_{-1} & | & \tilde{\rho}_1 & \cdots & \tilde{\rho}_n
\end{pmatrix}^\textnormal{T}.
\label{eq:F_matrix_2D}
\end{equation}

Suppose the matrix $F$ is invertible (by imposing the constraint $m=n$), then

\begin{equation}\label{eq_C(a)}
\bm{\tilde{\rho}'} \,=
\, F^{-1} (\bm{\hat{\rho}} - \tilde{\rho}_0\bm{1}).
\end{equation}

Substituting (\ref{eq_C(a)}) into (\ref{eq_5}), we finally get

\begin{equation}\label{eq:quadratic}
\tilde{\rho}_0^2 + [F^{-1}(\bm{\hat{\rho}} - \tilde{\rho}_0\bm{1})]^\textnormal{H} [F^{-1}(\bm{\hat{\rho}} - \tilde{\rho}_0\bm{1})] \,=\, (2\pi)^{-1} I.
\end{equation}

This is a quadratic equation in $\tilde{\rho}_0$, so we can solve it for $\tilde{\rho}_0$ first. Subsequently, we can solve \eqref{eq_C(a)} for $\bm{\tilde{\rho}'}$, so that all the Fourier coefficients $\bm{\tilde{\rho}}$ are found. This in turn allows for interpolation of the given spectral values $\hat{\rho}(\theta_1),\dots,\hat{\rho}(\theta_{2m})$ on the circle. Note that the symmetry relations \eqref{eq:plus_pi} and \eqref{eq:minus_k} can be exploited here. Alternatively, an adequate algorithm for solving the system \eqref{eq_C(a)} is the non-equispaced fast Fourier Transform (NFFT) of \cite{potts_steidl_tasche}. This algorithm and its so-called adjoint could also be used in the first step in an iterative scheme for solving the quadratic equation \eqref{eq:quadratic} for $\tilde{\rho}_0$. Performing the above-described procedure on a suitable number of concentric circles, we can potentially interpolate the complete spectrum $\hat{\rho}$ of $\rho$. In any case, from the interpolated spectral data, we can reconstruct the sought-after density $\rho$ from its spectrum $\hat{\rho}$ via a discretized inverse Fourier transform, cf.\ \eqref{eq:2d_Fourier_inverse}. This last step can be carried out using either a classical inverse FFT or the inverse NFFT of Potts et al., depending on the points at which the spectrum $\hat{\rho}$ is interpolated.

Before we show how the above method for SAXS-EM data fusion generalizes to the three-dimensional case, an important remark is in order. Of course, the above matrix $F$ will be invertible if and only if $m = n$ and if the projection angles $\theta_i$ are pairwise different from each other. If the number $m$ of EM projection angles is smaller than the approximation order $n$, than the Moore-Penrose pseudo inverse of the matrix $F$, or any other method for solving the linear system \eqref{eq:system} could be used. However, it clearly makes sense to choose the approximation order $n$ such that it matches $m$, so that the above-described method for SAXS-EM data fusion results in a quadratic and invertible matrix $F$. If some of the EM projection angles $\theta_i$ are very similar to each other, it could be numerically more stable to exclude the ``redundant'' angles, and to choose the approximation order $n$ accordingly. The over-determined case $m > n$ is neither of practical nor of theoretical interest for here the information of SAXS would theoretically already be included in the EM data.

Due to the geometric properties of the two-dimensional unit sphere, which are different from those of the three-dimensional unit sphere, the situation in the three-dimensional case is slightly more complicated than this, as we will see later. But the same conceptual framework applies.

\subsection{Numerical illustration - Fusion on a unit circle}
In this subsection, the basic principle behind SAXS-EM data fusion is illustrated numerically. To show the general concept, the experiment is performed in real space. But the calculation method can be implemented in Fourier space as well.

The following function on a unit circle is used as a testing model,
$$f(\theta) = (2 \sin^5\theta + \cos\theta +2\cos^3\theta)^2$$
In the experiment, it is supposed that values of the function are known when $\theta \in [30^\circ\ 35^\circ\ 38^\circ\ 40^\circ\ \\ 95^\circ\ 105^\circ\ 110^\circ\ 120^\circ]$, which is used as EM data.

The SAXS information can then be calculated as

\begin{equation}
I(1) = \int_{0}^{2\pi} |f(\theta)|^2 \mathrm{d} \theta \,
\label{eq_12}
\end{equation}

\begin{figure}[!ht]
\centering
\includegraphics[width = 10cm]{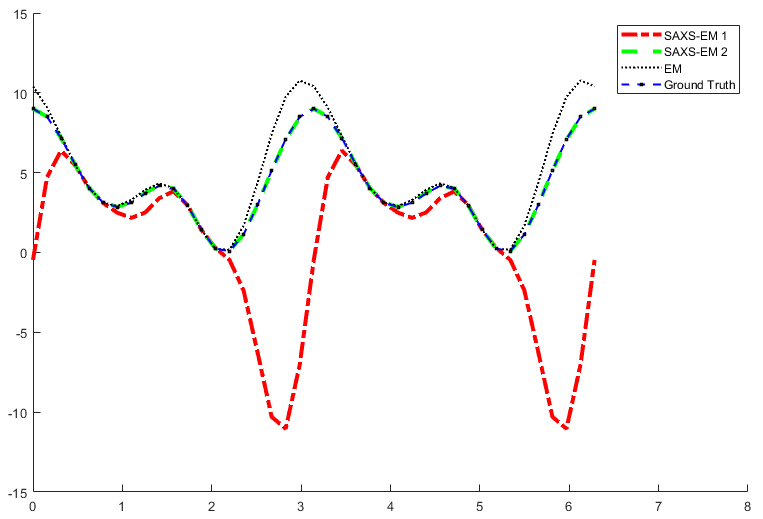}
\caption{Reconstruction results of EM and SAXS-EM comparing with the original function}
\label{Fig:1D_reconstruction}
\end{figure}

Both the EM reconstruction and SAXS-EM fusion reconstruction have been performed. Results are shown in Fig.~\ref{Fig:1D_reconstruction}, in which the two dashed lines (in red and green) are results from the SAXS-EM fusion, the black dotted line shows the EM result, and the blue dashed line with black dots is the original function.

\FloatBarrier

It can be seen from the figure, the SAXS-EM fusion technique provides two very different results. One fits the original function pretty well, but the other one goes far from the ground truth. The two result phenomenon can be explained since \eqref{eq:quadratic} is a quadratic equation solving for $\tilde{\rho}_0$, which yields two solutions. However, as shown in the figure, one of the two solutions is definitely wrong. 

In this 1D example, the value of the function can be used as the filter. Since the original function indicates the density of a model must be positive in each point, the red dash line can be easily eliminated. In the 2D and 3D cases, the minimum mean square error has been adopted to choose the solution which gives a more stable intensity value, $I(\rho)$. However, this  is a computationally intensive procedure and the origin of the two solutions obtained as well as the filtering steps will be explained in detail later in Section \ref{Roots}.

\subsection{The three-dimensional case}

In single-particle cryo EM, we have many copies of the same biomolecular structure (each called a ``particle'') oriented in many different ways. If $\rho : \IR^3 \rightarrow \IR_{\geq 0}$ is the density function of the particle as described in its own body-fixed frame, and if $R_i$ is the rotation matrix describing the orientation of the $i$\textsuperscript{th} particle relative to the laboratory frame, then the projection along the laboratory $z$-axis will be

\begin{align*}
\pi_i(x,y)
\,&\coloneqq\, \int_{-\infty}^{\infty} \rho(R_i^\textnormal{T} \bm{x}) \mathrm{d}z
\,=\, \int_{-\infty}^{\infty} \rho(x R_i^\textnormal{T} {\bm e}_1 + y R_i^\textnormal{T} {\bm e}_2 + z R_i^\textnormal{T} {\bm e}_3) \mathrm{d}z, 
\end{align*}

\noindent where $\bm{x} = (x,y,z)^\textnormal{T} \in \IR^3$, while ${\bm e}_1, {\bm e}_2, {\bm e}_3 \in \IR^3$ are respectively the canonical unit vectors in $x$-, $y$-, and $z$-direction.

This means that in the body-fixed frame of the $i$\textsuperscript{th} particle, the projection direction ${\bm n}_i$ is

$${\bm n}_i \,=\, R_i^T {\bm e}_3.$$

From the Fourier slice theorem, this then means that the two-dimensional Fourier transform $\hat{\pi}_i$ of the projection $\pi_i(x,y)$ is a slice of the three-dimensional Fourier transform,

$$\hat{\rho}(\bm{\omega}) \,=\, \int_{\mathbb{R}^3} \rho(\bm{x}) \mathrm{e}^{- \mathrm{i} \bm{\omega}^\textnormal{T} \bm{x}} \mathrm{d} \bm{x}$$
of the density function $\rho$ in the plane through the origin with normal ${\bm n}_i$. Explicitly, the Fourier slice theorem here reads

\begin{equation}
\hat{\pi}_i(\bm{\omega}) \,=\, \hat{\rho}(\omega_1 R_i^\textnormal{T} {\bm e}_1 + \omega_2 R_i^\textnormal{T} {\bm e}_2), \quad\quad \bm{\omega} \,=\, (\omega_1,\omega_2)^\textnormal{T} \in \IR^2.
\label{Fourierslice3d}
\end{equation}

The density function $\rho$ can be recovered from its Fourier transform $\hat{\rho}$ via the inverse Fourier transform,

\begin{equation}\label{eq:Fourier_inverse_3d}
\rho(\bm{x}) \,=\, (2\pi)^{-3} \int_{\mathbb{R}^3} \hat{\rho}(\bm{\omega}) \mathrm{e}^{\mathrm{i} \bm{\omega}^\textnormal{T} \bm{x}} \mathrm{d} \bm{\omega}.
\end{equation}

Let us assume that an existing EM reconstruction method has determined $R_1, \dots, R_n$, where $n$ is the total number of particles. Our goal is to fuse the experimental EM information for $i = 1,...,n $ given in the left-hand-side of \eqref{Fourierslice3d} with the SAXS experimental measurements. The SAXS data is the square average of the Fourier transform $\hat{\rho}$ over spheres centered at the origin, 

\begin{equation}
I(r) \,\coloneqq\, \int_{\mathbb{S}^2} |\hat{\rho}(r {\bm u})|^2 \mathrm{d}{\bm u} \,=\, \int_{\mathbb{S}^2} \hat{\rho}(r {\bm u}) \overline{\hat{\rho}(r {\bm u})} \mathrm{d}{\bm u},
\label{SAXS3d}
\end{equation}

\noindent where $\mathbb{S}^2$ is the two-dimensional unit sphere in $\IR^3$.

\begin{figure}[!h]
\centering
\includegraphics[width = 5cm]{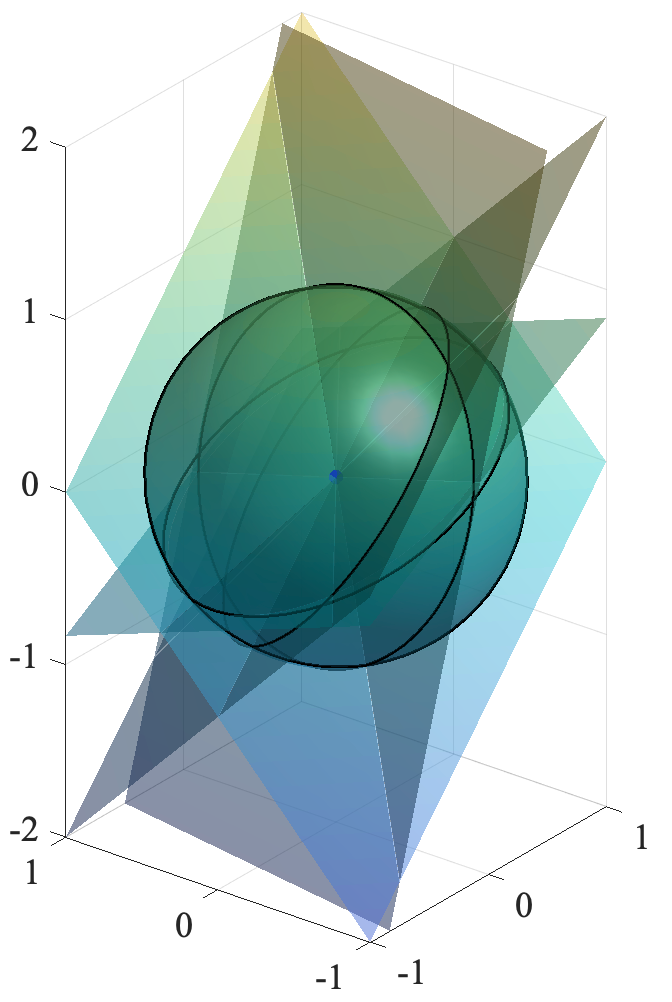}
\caption{Geometric situation in the three-dimensional case (conceptual plot).}
\label{Fig:example of 3D case}
\end{figure}

Our approach for SAXS-EM data fusion in the three-dimensional case will be a direct generalization of the two-dimensional case discussed in the previous section. We wish to reconstruct $\hat{\rho}$ 
by viewing the Fourier space as consisting of an infinite number of concentric spheres. On each of these spheres will be a certain number of so-called great circles, arising from the intersection of a plane containing the origin with the respective sphere, cf.\ Fig.\,\ref{Fig:example of 3D case}. The values of the spectrum $\hat{\rho}$ on these great circles are provided by EM data, cf.\ (\ref{Fourierslice3d}). The SAXS information (\ref{SAXS3d}) will serve to improve the reconstruction in the potentially large ``gaps'' between the great circles.

Apart from that, the main difference between the three-dimensional and the planar case is that instead of the classical Fourier basis $e^{\mathrm{i} k \theta}$, we will now employ spherical harmonics that we briefly review below.

\subsubsection{A brief review of spherical harmonics}

On the two-dimensional unit sphere $\mathbb{S}^2$ in $\IR^3$, let us introduce spherical coordinates $\theta$ and $\phi$ such that for any point $\bm{u} = (x,y,z)$ on the unit sphere $\mathbb{S}^2$ it holds

\begin{align*}
x \,&=\, \sin \theta \, \cos \phi, \\
y \,&=\, \sin \theta \, \sin \phi, \\
z \,&=\, \cos \theta.
\label{sphere11}
\end{align*}

The angle $\theta \in [0,\pi]$ is called the polar angle, while $\phi \in [0,2\pi)$ is referred to as the azimuthal angle. The integral of any integrable function $f : \mathbb{S}^2 \rightarrow \mathbb{C}$ can be computed via

$$\int_{\mathbb{S}^2} f({\bm u}) \mathrm{d}{\bm u} \,=\, \int_{0}^{\pi} \int_{0}^{2\pi} f(\theta,\phi) \mathrm{d}\phi \, \sin\theta \, \mathrm{d}\theta.$$

If a function $f$ is square-integrable over $\mathbb{S}^2$, \textit{i.\,e.},

$$\int_{\mathbb{S}^2} |f(\bm{u})|^2 \mathrm{d}\bm{u} \,<\,\infty,$$
then it can be expanded in terms of the spherical harmonics. These are defined as

\begin{equation*}
Y_{lm}(\theta,\phi) \,\coloneqq\, \sqrt{\frac{(2l+1)}{4\pi}\frac{(l-m)!}
{(l+m)!}} P_{l}^{m}(\cos \theta) \mathrm{e}^{\mathrm{i} m \phi}, \quad\quad l \in \{0, 1, 2, \dots\}, ~m \in \{-l,\dots,l\},
\label{sphereharm}
\end{equation*}

\noindent where
$$P^{m}_{l}(x) \,\coloneqq\, \frac{(-1)^m}{2^l l!} (1-x^2)^{m/2} \frac{\mathrm{d}^{l+m}}{\mathrm{d}x^{l+m}} (x^2-1)^l, \quad\quad x \in [-1,1],$$

\noindent is an associated Legenrde polynomial. This is possible because the spherical harmonics constitute an orthonormal basis for $L^2(\mathbb{S}^2)$, the set of all square-integrable functions on the sphere. Such an expansion reads as

\begin{equation}
f \,=\,
\sum_{l=0}^\infty \sum_{m=-l}^l \tilde{f}_{lm} Y_{lm},
\label{spherecompl}
\end{equation}

\noindent where
\begin{equation}
\tilde{f}_{lm} \,=\, \int_{\mathbb{S}^2} f(\bm{u}) \overline{Y_{lm}({\bm u})} \mathrm{d}{\bm u}
\label{eq:spherical_Fourier_coefficients}
\end{equation}
are the generalized, spherical Fourier coefficients of the function $f$. Equality in \eqref{spherecompl} holds in the $L^2$ sense, but in general not pointwise. However, as is commonly done, we can use a truncated version of the Fourier series as a pointwise approximation to the function $f$ of interest, which will be smooth in general.

Without going into too much detail, we note that the spherical harmonics have nice properties under rotation. Indeed, for a given rotation matrix $R$ we find that

\begin{equation*}
Y_{lm}(R^\textnormal{T} {\bm u}) \,=\,
\sum_{m'=-l}^l D_{m m'}^{(l)}(R) Y_{lm'}({\bm u}),
\end{equation*}

\noindent where $D_{m m'}^{(l)}$ are the well-studied Wigner-D functions (see \cite{biedenharn_louck, chirikjian2016harmonic} for example). In particular, this means that a rotated spherical harmonic can be written as a linear combination of spherical harmonics of the same degree.

\subsubsection{Data fusion}

As in the two-dimensional case, without loss of generality, let us confine ourselves to the unit sphere $\mathbb{S}^2$, \textit{i.\,e.}, the sphere with radius $r = 1$ in $\IR^3$. We can approximate the sought-after density spectrum $\hat{\rho}$ restricted to the unit sphere with a truncated spherical Fourier series,

\begin{equation*}
\hat{\rho}|_{\mathbb{S}^2} \,=\, \sum_{l=0}^{N-1} \sum_{m=-l}^l \tilde{\rho}_{lm} Y_{lm},
\end{equation*}

\noindent where $l \in \{1, 2, 3, \dots\}$ is the approximation order, and $\tilde{\rho}_{lm}$ are the spherical Fourier coefficients of $\hat{\rho}|_{\mathbb{S}^2}$, cf.\ \eqref{eq:spherical_Fourier_coefficients}.

The SAXS experiment gives

\begin{align}\label{eq_I_rho}
I \,\coloneqq\, I(1)
\,&=\, \int_{\mathbb{S}^2} \bigg(\sum_{l = 0}^{N-1} \sum_{m = -l}^l \tilde{\rho}_{lm} Y_{lm}(\bm{u})\bigg) \overline{\bigg(\sum_{l' = 0}^{N-1} \sum_{m' = -l'}^{l'} \tilde{\rho}_{l'm'} Y_{l'm'}(\bm{u}) \bigg)} \mathrm{d}{\bm u} \\\notag
\,&=\, \sum_{l = 0}^{N-1} \sum_{m = -l}^l \sum_{l' = 0}^{N-1} \sum_{m' = -l'}^{l'} \tilde{\rho}_{lm} \, \overline{\tilde{\rho}_{l'm'}} \int_{\mathbb{S}^2} Y_{lm}(\bm{u}) \overline{Y_{l'm'}(\bm{u})} \mathrm{d}{\bm u} \\\notag
\,&=\, \sum_{l = 0}^{N-1} \sum_{m = -l}^l \sum_{l' = 0}^{N-1} \sum_{m' = -l'}^{l'} \tilde{\rho}_{lm} \, \overline{\tilde{\rho}_{l'm'}} \, \delta_{ll'} \,  \delta_{mm'} \\\notag
\,&=\, \sum_{l = 0}^{N-1} \sum_{m = -l}^l |\tilde{\rho}_{lm}|^2.
\end{align}

Let us assume that the EM data $\hat{\rho}(\bm{u}_1),\dots,\hat{\rho}(\bm{u}_M)$ are given on the unit sphere at points $\bm{u}_1,\dots,\bm{u}_M \in \mathbb{S}^2$. Note that these points will lie on a certain number of great circles, cf.\ Fig.\,\ref{Fig:example of 3D case}. Analogous to \eqref{eq:plus_pi}, in the three-dimensional case we have that

\begin{equation}
\overline{\hat{\rho}(\bm{u})} \,=\, \hat{\rho}(-{\bm u}).
\label{eq:minus_u}
\end{equation}

Note that in spherical coordinates, $-{\bm u}$ parity transforms a point $\bm{u}$ with coordinates $(\theta,\phi)$ to $(\pi-\theta,\pi+\phi)$. From \eqref{eq:minus_u} and with the well-known relation $\overline{Y_{lm}} = (-1)^m {Y_{l,-m}}$, it can further be derived that (cf.\ \eqref{eq:minus_k})

\begin{equation}\label{eq_C_hat}
\tilde{\rho}_{lm} \,=\, (-1)^{l+m} \, \overline{\tilde{\rho}_{l,-m}}.
\end{equation}

Let us introduce the EM-data vector

\begin{equation*}
\bm{\hat{\rho}} \,\coloneqq\, (\hat{\rho}(\bm{u}_1),\dots,\hat{\rho}(\bm{u}_M))^\textnormal{T},
\end{equation*}

\noindent as well as the sought-after frequency vector

\begin{equation*}
\bm{\tilde{\rho}} \,\coloneqq\, (\tilde{\rho}_{0,0}\,|\,\tilde{\rho}_{1,-1},\,\tilde{\rho}_{1,0},\,\tilde{\rho}_{1,1}\,|\,\tilde{\rho}_{2,-2},\dots,\tilde{\rho}_{2,2}\,|\,\dots\,|\,\tilde{\rho}_{N-1,1-N},\dots,\tilde{\rho}_{N-1,N-1})^\textnormal{T}.
\end{equation*}

The SAXS data \eqref{eq_I_rho} thus attain the form

\begin{equation}\label{eq:SAXS_3d}
I \,=\, \bm{\tilde{\rho}}^\textnormal{H} \bm{\tilde{\rho}}.
\end{equation}

Now we can state the EM reconstruction problem in matrix-vector notation as
\begin{equation}
\bm{\hat{\rho}} = B \bm{\tilde{\rho}}
\label{eq_matrix}
\end{equation}

\noindent where
$$B = \begin{pmatrix}
\begin{array}{c|ccc|c|c}
Y_{0,0}(\bm{u}_1) & Y_{1,-1}(\bm{u}_1) & Y_{1,0}(\bm{u}_1) & Y_{1,1}(\bm{u}_1) & \ldots & Y_{N-1,N-1}(\bm{u}_1) \\
Y_{0,0}(\bm{u}_2) & Y_{1,-1}(\bm{u}_2) & Y_{1,0}(\bm{u}_2) & Y_{1,1}(\bm{u}_2) & \ldots & Y_{N-1,N-1}(\bm{u}_2) \\
\vdots & \vdots & \vdots & \vdots & & \vdots\\
Y_{0,0}(\bm{u}_M) & Y_{1,-1}(\bm{u}_M) & Y_{1,0}(\bm{u}_M) & Y_{1,1}(\bm{u}_M) & \ldots & Y_{N-1,N-1}(\bm{u}_M)
\end{array}
\end{pmatrix}  \in\, \mathbb{C}^{M \times N^2}.$$

When $l = m = 0$, from (\ref{eq_C_hat}), we have $\tilde{\rho}_{0,0} = \overline{\tilde{\rho}_{0,0}}$ and so $\tilde{\rho}_{0,0}$ is real. Bringing this unknown to the left-hand side while noting that $Y_{0,0} = (4 \pi)^{-1/2}$, the system (\ref{eq_matrix}) transforms into

\begin{equation}\label{eq_matrix_2}
\bm{\hat{\rho}}- \frac{\tilde{\rho}_{0,0}}{\sqrt{4\pi}}
\bm{1} \,=\, F\bm{\tilde{\rho}'}
\end{equation}

\noindent where $\bm{1}$ is the vector of length $M$ containing only ones,

$$
F\,=\,
\begin{pmatrix}
\begin{array}{ccc|c|c}
Y_{1,-1}(\bm{u}_1) & Y_{1,0}(\bm{u}_1) & Y_{1,1}(\bm{u}_1) & \ldots & Y_{N-1,N-1}(\bm{u}_1) \\
Y_{1,-1}(\bm{u}_2) & Y_{1,0}(\bm{u}_2) & Y_{1,1}(\bm{u}_2) & \ldots & Y_{N-1,N-1}(\bm{u}_2) \\
\vdots & \vdots & \vdots & & \vdots\\
Y_{1,-1}(\bm{u}_M) & Y_{1,0}(\bm{u}_M) & Y_{1,1}(\bm{u}_M) & \ldots & Y_{N-1,N-1}(\bm{u}_M)
\end{array}
\end{pmatrix} \in \, \mathbb{C}^{M \times (N^2-1)},
$$

$$
\bm{\tilde{\rho}'}\,=\,
\begin{pmatrix}
\tilde{\rho}_{1,-1}\ &\tilde{\rho}_{1,0} \ &\tilde{\rho}_{1,1} \ &| \cdots | \ & \tilde{\rho}_{N-1,N-1}
\end{pmatrix}^\textnormal{T}.
$$

With the Moore-Penrose pseudo inverse $F^\dagger$ of $F$, the system \eqref{eq_matrix_2} can be solved as

\begin{equation}\label{eq:Fourier_system_3d}
\bm{\tilde{\rho}'} \,=\, F^\dagger (\bm{\hat{\rho}} - \tilde{\rho}_{0,0} \bm{1} / \sqrt{4\pi}).
\end{equation}

Inserting this equation into \eqref{eq:SAXS_3d} then yields

\begin{equation}\label{eq:quadratic_3d}
\tilde{\rho}_{0,0}^2 + [F^\dagger (\bm{\hat{\rho}} - \tilde{\rho}_{0,0} \bm{1} / \sqrt{4\pi})]^\textnormal{H} [F^\dagger (\bm{\hat{\rho}} - \tilde{\rho}_{0,0} \bm{1} / \sqrt{4\pi})] \,=\, I.
\end{equation}

As in the two-dimensional case, this is a quadratic equation with respect to the first unknown Fourier coefficient $\tilde{\rho}_{0,0}$. After solving this quadratic equation, the remaining Fourier coefficients $\bm{\tilde{\rho}'}$ can be determined by solving the linear system \eqref{eq:Fourier_system_3d} subject to \eqref{eq:minus_u} and \eqref{eq_C_hat}, so that all the Fourier coefficients $\bm{\tilde{\rho}}$ are found by inverting \eqref{eq_matrix}, which we do via the pseduo inverse based on the SVD. An alternative tool for this task is the well-known fast non-equispaced spherical Fourier transform of \citet{kunis_potts}. This fast algorithm and its adjoint can also be used in the first step, in which the quadratic equation \eqref{eq:quadratic_3d} would then be solved for $\tilde{\rho}_{0,0}$ with a suitable iterative technique. Once the spherical Fourier coefficients $\bm{\tilde{\rho}}$ are determined, the spectral EM data $\bm{\hat{\rho}}$ can be interpolated on the sphere. After this is done on a suitable number of concentric spheres, the spectrum $\hat{\rho}$ of $\rho$ can potentially be interpolated in $\mathbb{R}^3$. The sought-after density function $\rho$ can now be reconstructed from the interpolated spectral data via a discretized inverse Fourier transform, cf.\ \eqref{eq:Fourier_inverse_3d}.

To close this section, let us comment on the relation between the number $M$ of EM data points considered on the unit sphere $\mathbb{S}^2$ above and the approximation order $N$. In the two-dimensional case, we saw that this relation was very clear: As long as the different projection angles $\theta_i$ in EM are not too similar to each other, the approximation order $m$ should be chosen to be equal to the number of projection angles $n$, for this will result in a quadratic and invertible Fourier matrix $F$ in \eqref{eq:system}. As mentioned above, the situation in the three-dimensional case is slightly more complicated than that. In particular, it would in general not be sufficient to choose the approximation order $N$ equal to $\sqrt{M}$, although this would mean that we had as many data points as spherical Fourier coefficients that we wish to reconstruct ($N^2$). The reason for this is that the EM data will always lie on great circles (again, cf.\ Fig.\,\ref{Fig:example of 3D case}), which is an inherent problem in EM imaging if reconstruction is performed using concentric spheres. Indeed, it is well known from constructive approximation theory on the sphere that $N^2$ pairwise different sample points on \emph{one} great circle are not sufficient for reconstructing a spherical polynomial of degree $N-1$ with $N^2$ Fourier coefficients (see \cite{freeden_gervens_schreiner}, for instance). The situation is slightly relaxed by the fact that there will be more than one great circle in practical applications. The number of different projection angles, however, can be quite limited, and there might be large gaps between the projection planes, where information is missing. This makes including information available from SAXS even more desirable.

\subsubsection{Numerical illustration - Fusion on a unit sphere}\label{Gaussian_Sph}

This numerical experiment shows the difference between the EM reconstruction and the SAXS-EM reconstruction. The aim is to illustrate that when additional data from SAXS experiments is supplemented, the reconstruction is better for the three-dimensional case. The calculation is performed in Fourier space directly.

The 3D function used in this experiment is
$$f(\bm{x}) = e^{-12(\bm{x} - \bm{x}_{o1})^2} - e^{-12(\bm{x} - \bm{x}_{o2})^2}$$

\noindent where $\bm{x}_{o1} = \begin{bmatrix} 1/\sqrt{2} & 0 & 1/\sqrt{2} \end{bmatrix}$, $\bm{x}_{o2} = \begin{bmatrix} 0 & -1 & 0 \end{bmatrix}$. 

\begin{figure}[t]
\centering
\includegraphics[width = 5cm]{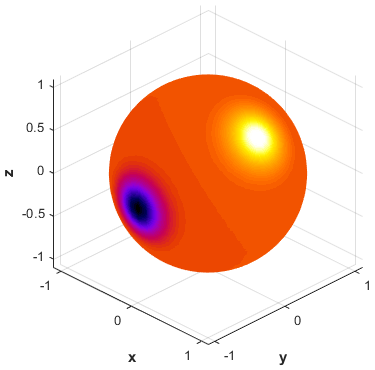}
\caption{Geometric situation in the three-dimensional case (conceptual plot).}
\label{Fig:3D_sphere_ground_truth}
\end{figure}

The value of the function on a unit sphere ($r = 1$) is shown in Fig.~\ref{Fig:3D_sphere_ground_truth}. In the numerical experiment, 20 projections of the function were randomly generated. On each projection, which is a unit sphere, 50 evenly distributed points are selected with known values. This constitutes the EM data.

\FloatBarrier

The roots that the quadratic equation in (\ref{eq:quadratic_3d}) is expected to generate can be compared with an analytical solution calculated using (\ref{eq:spherical_Fourier_coefficients}),
$$\tilde{\rho}_{0,0} = \frac{\pi}{\sqrt{4\pi}}(\frac{1-e^{-4 \times 12}}{12} - \frac{1-e^{-4 \times 12}}{12})$$

\noindent and hence, $\tilde{\rho}_{0,0} = 0$.

\begin{figure}[!h]
\center
\subfigure[]{\includegraphics[width=5cm]{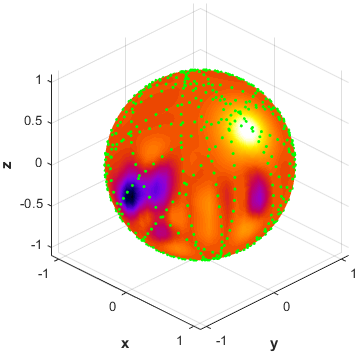}\label{Fig:3D_sphere_SAXS-EM_reconstruction}}
\hfil
\subfigure[]{\includegraphics[width=5cm]{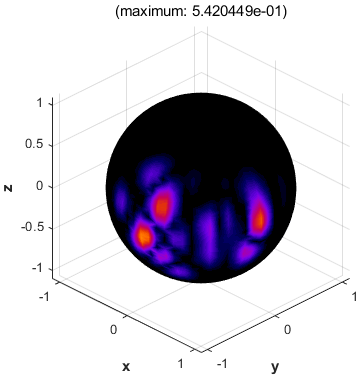}\label{Fig:3D_sphere_SAXS-EM_error}}
\caption{Result of SAXS-EM reconstruction and its absolute error}
\label{Fig:3D_sphere_SAXS-EM}
\end{figure}

\begin{figure}[!h]
\center
\subfigure[]{\includegraphics[width=5cm]{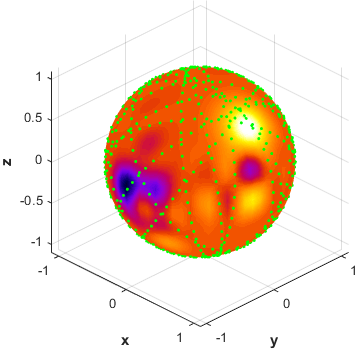}\label{Fig:3D_sphere_EM_reconstruction}}
\hfil
\subfigure[]{\includegraphics[width=5cm]{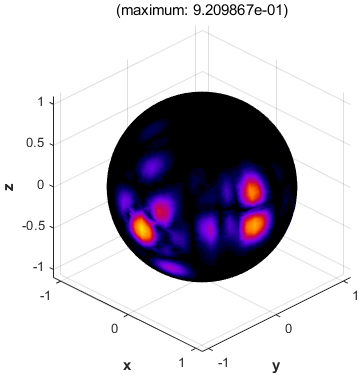}\label{Fig:3D_sphere_EM_error}}
\caption{Result of EM reconstruction and its absolute error}
\label{Fig:3D_sphere_EM}
\end{figure}

Figure~\ref{Fig:3D_sphere_EM_reconstruction} shows the reconstruction result using only the EM data. Qualitatively the reconstruction creates fictitious spots on the sphere and does not reproduce the peaks of the Gaussian correctly. This is reflected in the absolute error plot adjoining it. The reconstruction result from SAXS and EM fusion and the corresponding absolute error are shown in Fig.~\ref{Fig:3D_sphere_SAXS-EM}.

\FloatBarrier
\FloatBarrier

By comparing Fig.~\ref{Fig:3D_sphere_EM_error} and Fig.~\ref{Fig:3D_sphere_SAXS-EM_error}, the reconstruction error has been reduced $50\%$ by fusing SAXS data. Furthermore, it is worth to note that the coarser the EM data sampling, the greater the reconstruction error. In such instances, the role of SAXS data becomes more valuable.
\section{Structure and Selection of the Quadratic Roots}\label{Roots}
In this section, we describe in greater detail the origin, underlying structure as well as the method of selection employed to choose between the roots of the quadratic equations in (\ref{eq:quadratic}) and (\ref{eq:quadratic_3d}). 

Let $a$ represent the coefficient of the zero-order term in the Fourier or spherical harmonic expansion (\textit{i.\,e.} $\rho_0$ and $\rho_{0,0}$ respectively). The two roots of the quadratic equation, $a^1$ and $a^2$, and they can be used to derive two different sets of $\bm{\tilde{\rho}}$ from (\ref{eq_C(a)}) and (\ref{eq:Fourier_system_3d}): $\bm{\tilde{\rho}}^1$ and $\bm{\tilde{\rho}}^2$ . From the intensity constraint ((\ref{eq_I_rho}) for 2D and (\ref{eq:SAXS_3d}) for 3D), we observe that for any $\bm{\tilde{\rho}}$ that satisfies the constraint, a transformed coefficient vector, $U\bm{\tilde{\rho}}$ where $U$ is a unitary matrix (\textit{i.\,e.} $UU^H = I$) also satisfies the constraint. Hence, we can write $\bm{\tilde{\rho}}^2 = U\bm{\tilde{\rho}}^1$. We now concentrate on the 2D case, although a similar analysis can be performed for the 3D scenario as well.
\subsection{Origin of the roots in 2D}
We let $2n + 1$ denote the dimension of the $\bm{\tilde{\rho}}$ vector (the superscript will henceforth be omitted and the relationship between the two solutions be made clear through the unitary matrix) and let $m = 2n$ be the dimension of the EM data, $\bm{\hat{\rho}}$. As we proceed to explain, the difference of $1$ between the two dimensions leaves the system under-determined and therefore allows the intensity constraint to close the system of equations. If the difference between the two dimensions were larger, one would require the Moore-Penrose pseudo inverse to approximate the solutions (as is performed in the three-dimensional scenario). We proceed by making some observations regarding the under-determined nature of the system of equations.
Since $\bm{\tilde{\rho}}$ and $U\bm{\tilde{\rho}}$ both satisfy equation (\ref{eq:eq_matrix}),

\begin{equation}\label{eq:spur_tworoot}
\bm{\hat{\rho}} = A \bm{\tilde{\rho}} = AU\bm{\tilde{\rho}}.
\end{equation}

This implies that,
\begin{equation}\label{eq:spur_null}
A(U - I)\bm{\tilde{\rho}} = \bm{0}.
\end{equation}

Hence, the matrix $U - I$ transforms a non-zero $\bm{\tilde{\rho}}$ such that it finds itself in the null space of $A$. This result implies that $A$ has linearly independent rows but linearly dependent columns. This is a consequence of finite discretization: EM data that is known at some points can allow for more than one interpolation of the Fourier transform between the data points. The quadratic nature of the intensity constraint narrows the space of admissible solutions (that satisfy the discrete set of equations) to two. Roots generated after gradual refinement of the EM data with more slices can be expected to converge towards the true root. A related method to select the true root without needing refined EM data would be to selectively block data obtained at various angles and evaluate the roots. The spurious root would be expected to fluctuate significantly whereas the true root would be fairly stable throughout this blocking and calculation procedure. However such a method is also computationally intensive especially in two or three dimensional cases. In such scenarios other techniques of root selection would be important, and these would be highlighted later.

\subsection{Origin of the roots in 3D}
A similar procedure can be performed for 3D although the situation is more complicated. We let $N^2$ denote the dimension of the $\bm{\tilde{\rho}}$ vector and let $M$ be the dimension of the EM data, $\bm{\hat{\rho}}$. Again the set of equations would be under-determined since due to the interpolation conditions on a sphere, $\sqrt{N}$ data points are not sufficient to interpolate an $N-1$ degree spherical harmonic on the spherical surface and we would obtain an interpolation of lower ``effective" order. Recognizing this fact, we yet again have a system that allows two roots to be connected by a unitary transformation such that the difference between the two roots falls in the null space of the matrix, $B$ in (\ref{eq_matrix}). Note that the fundamental assumption is that the system of equations is originally under-determined and allows for the (approximate) evaluation of a root. As mentioned earlier, over-determined systems would not be of practical interest.

\subsection{Types of roots and relationship with intensity}
The quadratic equation (\ref{eq:quadratic}) in the two-dimensional problem can be expanded as,

\begin{equation}\label{eq:quadratic_2Dsimple}
  \left(1+\bm{1}^HM\bm{1}\right)\tilde{\rho}_0^2-2{\hat{\rho}}^HM\bm{1}\tilde{\rho}_0+{\hat{\rho}}^HM\hat{\rho}-\frac{I}{2\pi}=0
\end{equation}

\noindent where $M = (FF^{H})^{-1}$ is a Hermitian matrix that contains information about the distribution of the EM sampled points. The roots of this equation are,

\begin{equation}\label{eq:Quadroots}
  \tilde{\rho}_0= \frac{\hat{\rho}^HM\bm{1}\pm\sqrt{(1+\bm{1}^HM\bm{1})[\frac{I}{2\pi}-\hat{\rho}^H(M-\frac{M\bm{1}\bm{1}^HM}{1+\bm{1}^HM\bm{1}})\hat{\rho}]}}{1+\bm{1}^HM\bm{1}}.
\end{equation}

We also observe that the intensity only appears in the discriminant, which is proportional to the term under the square root in (\ref{eq:Quadroots}). In fact, since $\bm{1}^HM\bm{1} \geq 0$ using the positive semi-definite property of the Hermitian matrix, $M$, the nature of roots is solely controlled by the sign of $[\frac{I}{2\pi}-\hat{\rho}^H(M-\frac{M\bm{1}\bm{1}^HM}{1+\bm{1}^HM\bm{1}})\hat{\rho}]$. 
\newpage
Now, we define $I^{*}$ as the following expression,

\begin{equation}\label{eq:exact_intensity}
  I^{*}={2\pi}\hat{\rho}^H\left(M-\frac{M\bm{1}\bm{1}^HM}{1+\bm{1}^HM\bm{1}}\right)\hat{\rho}.
\end{equation}

Here, $I^{*}$ can be interpreted as an intensity since it has a similar form as the original intensity constraint in ($\ref{eq:SAXS_3d}$). Moreover, it will be shown that $I^{*}$ is in fact the lowest possible intensity that the EM data can admit, and is associated with the Fourier interpolation with the lowest norm that is used to fit the EM data. All values of $I \neq I^{*}$ will allow for two different roots to exist. If $I < I^{*}$ then the roots are complex whereas when $I \geq I^{*}$ the roots are real. In fact, the difference between the two roots would be proportional to $\sqrt{I - I^{*}}$. The constant of proportionality would be related to the sampling of angles in the EM data since the scalar $1+\bm{1}^HM\bm{1}$ is related to the angular spacing between various sampling points (and will have a considerably simple structure if the samples are uniformly spaced). 
A similar form of $I^{*}$ exists for the three-dimensional problem where we have,
\begin{equation}\label{eq:exact_intensity_3d}
  I^{*}=\hat{\rho}^H\left(M-\frac{M\bm{1}\bm{1}^HM}{4\pi+\bm{1}^HM\bm{1}}\right)\hat{\rho}.
\end{equation}
\noindent where $M = (FF^{H})^{\dagger}$ and is now expressed in terms of the pseudo-inverse.
\paragraph{The special case of complex conjugate roots:}
Since the matrix $M = (FF^{H})^{-1}$ is Hermitian, the coefficients of the quadratic equation (\ref{eq:quadratic}) are real (a similar conclusion can also be drawn for the three-dimensional case in (\ref{eq:quadratic_3d})). The three possible solutions of $\tilde{\rho}_0$ would be repeated real roots, distinct real roots and a complex conjugate pair. Here, we make a remark on the significance of the complex conjugate pair of roots.
A complex conjugate pair of roots would imply that $\overline{\bm{\tilde{\rho}}^1}=\bm{\tilde{\rho}}^2$. Relating the two vectors by the unitary matrix transformation, we get,

\begin{equation}\label{eq:spur_unitarycmplx1}
\overline{\bm{\tilde{\rho}}^1} = U\bm{\tilde{\rho}}^1.
\end{equation}

Multiplying both sides by the inverse of the Hermitian matrix, $U^H$ and taking the conjugate of both sides, 
\begin{equation}\label{eq:spur_unitarycmplx2}
\overline{\bm{\tilde{\rho}}^1} = U^T\bm{\tilde{\rho}}^1.
\end{equation}

Equating the right hand side of (\ref{eq:spur_unitarycmplx1}) and (\ref{eq:spur_unitarycmplx2}), we obtain $U = U^T$ for a non-zero $\bm{\tilde{\rho}}^1$. We also point out that a symmetric unitary matrix has many interesting properties, especially under the decomposition, $U = A + iB$ for symmetric real matrices, $A$ and $B$. We observe from $UU^{H} = I$ that $A^2 + B^2 = I$ and $AB = BA$. Particularly, if we define $Y = B + iA$, then $YU = UY = iI$. 
However, the existence of complex roots violates the important symmetry condition in equation (\ref{eq:minus_k}) and its corresponding form in the three-dimensional case (\ref{eq_C_hat}). A complex root suggests that no real value of $\tilde{\rho}_0$ can satisfy both the intensity as well as the EM data fitting constraint in the given set of EM data. Complex roots (since it appears in pairs) are yet again symptoms of discretisation introduced by the finite set of EM data and will disappear with finer sampling. A simple numerical technique to deal with complex roots would be to find the closest real root to approximate a given complex root that in this case would be the real component of the complex root since it corresponds to the minimum point of the parabola representing the quadratic expression. However it must be noted that by removing the imaginary component of the root, one also removes the information provided by the SAXS fusion data (since the intensity appears in the discriminant, which finds itself only in the imaginary component of the complex root). In such scenarios, the SAXS-EM fusion would degenerate to a reconstruction from EM data alone. The situations when the SAXS information fusion degenerates to an EM reconstruction will be explained in the following section in greater detail.

\subsection{Lower bound for SAXS intensity in SAXS-EM fusion}\label{LowerBound}
In this subsection, we aim to establish a connection between the results obtained from the SAXS-EM fusion and those obtained from EM data alone. By doing so, it is possible to shed light on the characteristics of the roots obtained in (\ref{eq:Quadroots}). 

Let a vector, $\bm{\tilde{\rho}}$, be evaluated in the EM experiment such that,

\begin{equation}\label{eq:EM_rho}
    \bm{\tilde{\rho}} = C^{\dagger}\bm{\hat{\rho}}.
\end{equation}

Here, $C$ is a matrix that corresponds to $A$ (\ref{eq:eq_matrix}) in the two-dimensional case  or $B$ (\ref{eq_matrix}) in the three-dimensional case. It is important to point out that since $C^{\dagger}$ is usually under-determined, (\ref{eq:EM_rho}) would have an infinite number of solutions and without any other constraint, the Moore-Penrose pseudo inverse would select the solution, $\bm{\tilde{\rho}}$ with the lowest Euclidean norm. The value of the norm is proportional to the intensity obtained from the EM calculations, \textit{i.\,e.},
\begin{equation}\label{eq:EM_I}
    I_{EM} = 2\pi\bm{\tilde{\rho}}^H\bm{\tilde{\rho}}.
\end{equation}
Substituting (\ref{eq:EM_rho}) in (\ref{eq:EM_I}) and expressing $C^{\dagger}$ as $C^H(CC^H)^{-1}$ (which is true for under-determined systems), we obtain,
\begin{equation}\label{eq:EM_I_exp}
    I_{EM} = 2\pi\bm{\hat{\rho}}^H(CC^H)^{-1}\bm{\hat{\rho}}
\end{equation}
using the Hermitian property of $CC^H$. The matrix, $CC^H$, can be related to the $F$ matrix defined in equations (\ref{eq:F_matrix_2D}) and (\ref{eq_matrix_2}) as $CC^H = FF^H + \bm{1}\bm{1}^H$ where $\bm{1}$ is a column vector of ones. The inverse of $CC^H$ appears in (\ref{eq:EM_I_exp}) and the formula derived by \cite{Sherman1950} allows us to express,
\begin{align}\label{eq:InvertSum}
\nonumber (CC^H)^{-1} &= (FF^H + \bm{1}\bm{1}^H)^{-1} \\
            &= (FF^H)^{-1} - \frac{(FF^H)^{-1}\bm{1}\bm{1}^H(FF^H)^{-1}}{1 + \textrm{tr}(\bm{1}^H(FF^H)^{-1}\bm{1})}.
\end{align}

Here $\textrm{tr}$ represents the trace operator. Substituting $M = (FF^H)^{-1}$ in (\ref{eq:InvertSum}) and using the result in (\ref{eq:EM_I_exp}), we see that,
\begin{equation}\label{eq:Equivalence}
I_{EM} = I^{*}
\end{equation}
\noindent where $I^{*}$ was defined in (\ref{eq:exact_intensity_3d}).

The result has important implications in understanding the usefulness of the SAXS intensity information in the fusion procedure. Firstly, we emphasize that $I_{EM}$ sets a minimum bound to the intensity value that can be used to set a constraint on a given set of EM data. When the SAXS intensity information is greater than this value, \textit{i.\,e.} $I > I_{EM}$, one obtains two real solutions. As shown earlier, these solutions are related by a unitary transformation and geometrically they can be visualized as two points on a constant intensity sphere in $\bm{\tilde{\rho}}$ space that can be related through reflections or rotations. When $I = I_{EM}$ the SAXS-EM fusion degenerates to an EM reconstruction alone and the SAXS intensity information adds no further information to the EM data. Geometrically, the solution space collapses to a point and the only unitary transformation that can exist is the identity transformation. When $I < I_{EM}$, it is not possible to fuse the SAXS information with the given set of EM data without violating some of the constraints (in this case, the symmetry conditions). In practical usage, the imaginary term is dropped such that we yet again obtain the situation where $I = I_{EM}$ and the SAXS-EM fusion degenerates to an EM reconstruction. Hence, SAXS intensity information can only add additional information to the results from the EM experiment when $I > I_{EM}$. 

We also make a final remark that when the EM data is sampled over concentric circles or spheres, the lower bound of intensity applies separately for each radius. In overall, it is reasonable to expect that for some, but not necessarily all, radii the SAXS-EM fusion would degenerate to an EM reconstruction. Additionally, this result as of yet makes no prediction on whether the SAXS-EM reconstruction is ``better" than the EM reconstruction. This would depend on other factors as well such as sampling, form of function to be reconstructed and order of the harmonic interpolation and on the conditioning of the relevant matrices.

\subsection{Root Selection Procedure}
\begin{figure}[!h]
\centering
\includegraphics[width = 9cm]{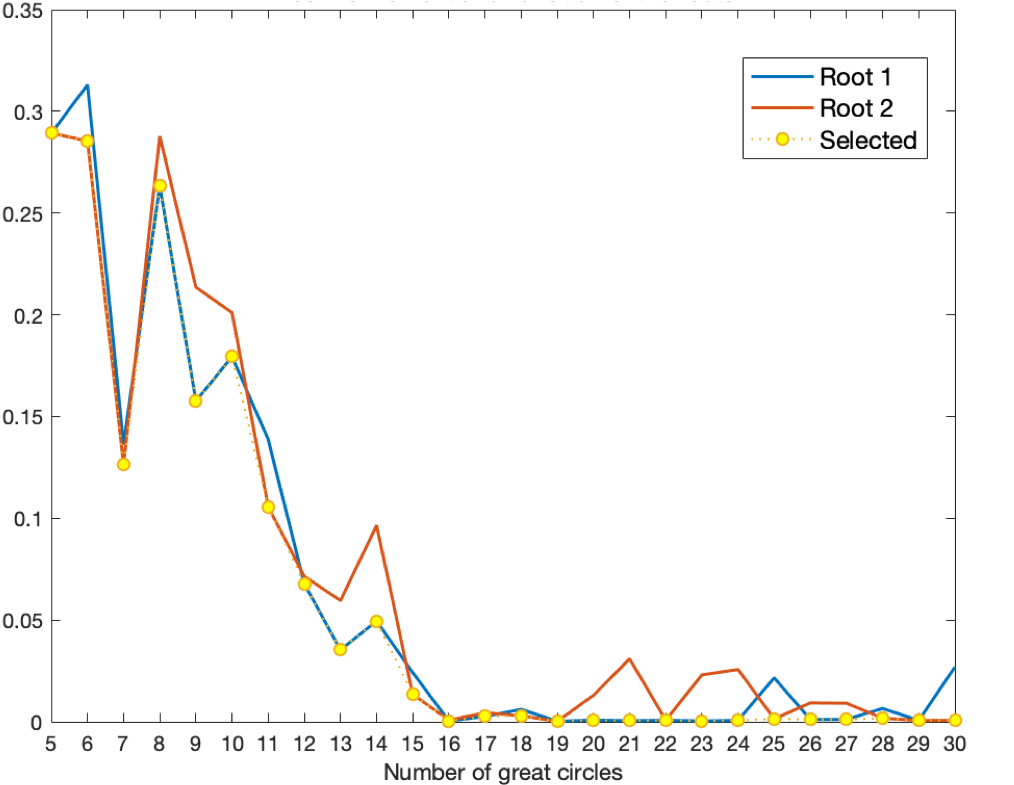}
\caption{Convergence of the absolute magnitudes of the roots with increasing number of sampled great circles.}
\label{Fig:Root_Convergence}
\end{figure}

In the development of the numerical algorithm to implement SAXS-EM fusion, the important question of how to select the ``correct" root needs to be answered. For the case of complex and repeated roots, this is not a problem since in both cases the real part of the two roots are the same. However, for the case of real but distinct roots, an additional root selection method needs to be devised. The underlying concept behind the creation of such a method is to observe that the spurious root would be more highly dependent on discretiszation parameters such as the sample points and their spacing. On the other hand, the true root would tend to be stable as the mesh is refined or the sample points modified. For instance, consider the Fig.~\ref{Fig:Root_Convergence} that shows the variations of the two roots with increasing refinement of samples of EM data for the problem in Section \ref{Gaussian_Sph}: roughly, the higher the number of great circles sampled, the lesser the difference in the absolute values of the two roots. Both roots converge to the analytic solution of zero as the number of great circles increases. However, the convergence is not monotonic since the great circles are randomly chosen for each step of sample refinement. Hence, the fluctuations depend on the sensitivity of a given root to the set of sample points; the greater the fluctuation the greater the likelihood that a given root can be considered as ``spurious". The yellow markers indicate the root that is ultimately selected by the algorithm to perform the SAXS-EM fusion for the given set of data. In this specific case, root 1 appears to experience fluctuations of lower amplitude and is selected in preference to root 2 for most part. The selection was carried out by a direct summation method, which will be explained further in the subsequent paragraphs. 

\FloatBarrier

The problem of selecting the correct root is complicated by the fact that with the discrete set of EM data, one does not usually have \textit{a priori} knowledge of the final reconstructed function. In some cases, it may be possible to assume that the final reconstruction should be fairly smooth and hence spurious Fourier interpolations that fluctuate greatly over the domain of EM data can be eliminated. This can be done by integrating (\ref{eq:tilde_rho}) or (\ref{eq:spherical_Fourier_coefficients}) numerically, using the EM data, to obtain an approximate value for $\rho_{0}$ or $\rho_{0,0}$ respectively. This value forms a guess or a ``selector" and would then be compared with the roots of the quadratic equation and the value with a lower absolute difference is chosen. This working assumption would then be that the numerical integration of the original EM data to obtain the zeroth order Fourier coefficient would be similar to the numerical integration of the interpolated data and this may be true, roughly, if the fluctuations have a period greater than the average spacing between the discrete EM data samples.

There are many ways to carry out this integration. One way would be to interpolate the original EM data (using a polynomial for instance) over the sphere of interest and then evaluate the integrals in (\ref{eq:tilde_rho}) and (\ref{eq:spherical_Fourier_coefficients}) through a quadrature rule. Another method would be to approximate the integral as a summation over the EM data as the following in three-dimensions,

\begin{equation}\label{eq:DirectSumApprox3D}
    \tilde{\rho}_{0,0} \approx \frac{1}{\sqrt{4\pi}} \sum_{i=1}^{2m}\hat{\rho}(\theta_i,\phi_i)\sin\theta_i\Delta\theta\Delta\phi
\end{equation}
and as,
\begin{equation}\label{eq:DirectSumApprox2D}
    \tilde{\rho}_{0} \approx \frac{1}{2\pi}\sum_{i=1}^{2m}\hat{\rho}(\theta_i)\Delta\theta.
\end{equation}
for the two-dimensional case.

Here, there are $2m$ points in the EM data set and $\Delta\theta$ and $\Delta\phi$ represent an average grid spacing in the polar and azimuthal directions respectively. The following plot presents the convergence rates of the roots along with that of the selector, again applied to the example in Section \ref{Gaussian_Sph}.

\begin{figure}[!h]
\centering
\includegraphics[width = 9cm]{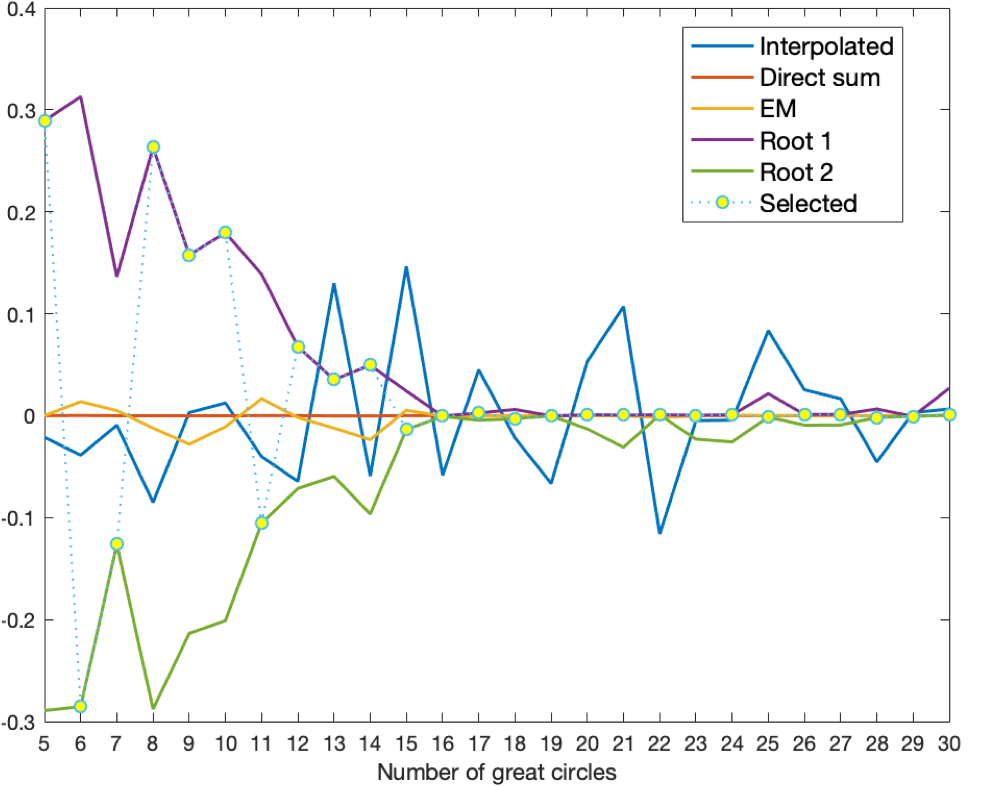}
\caption{The variation of the real parts of the roots as well as that of the root selectors (evaluated by direct summation or interpolation) with increasing number of sampled great circles. The imaginary parts of the roots are of order $10^{-15}$ and have been neglected.}
\label{Fig:Root_Convergence_2}
\end{figure}
\FloatBarrier

From the figure it is clear that interpolating the EM data over the sphere followed by integration to obtain the zeroth order Fourier series coefficient is worse as a selector as opposed to approximating the root through a direct sum method in (\ref{eq:DirectSumApprox3D}) or (\ref{eq:DirectSumApprox2D}). The analytic solution for the root is zero for this particular example and it can be seen that the selector evaluated through an interpolation followed by numerical integration converges too slowly with sample refinement. The roots are therefore selected through (\ref{eq:DirectSumApprox3D}) and plotted as the green circles. 
The root evaluated directly from the EM data is also plotted; this corresponds to the value of $\rho_{0,0}$ that results in a unique real solution for (\ref{eq:quadratic_3d}). Since we know from Section \ref{LowerBound} that this solution is the average of the two roots evaluated from SAXS-EM fusion, the EM root would not select one root in the preference of another as both SAXS-EM fusion roots differ by the same absolute magnitude from it; hence, the EM solution forms an unbiased reference to compare various root selection strategies against each other. 

Another approach to selecting roots is to repeatedly block angles of EM information, one data point at a time, and calculate the roots for the angles that remain. This is repeated for all angles in the set of EM data and the root whose value is nearly stable in this blocking and calculating procedure is selected. Although this does allow for a good reconstruction, as shown in Section \ref{SumCo-originGauss}, it is a computationally intensive procedure and was replaced by (\ref{eq:DirectSumApprox3D}) and (\ref{eq:DirectSumApprox2D}) for more complicated shapes such as the ``smiley" and ``minion" tested later.

\section{Numerical Experiments}
\label{sec:numerical_experiments}

\subsection{Sum of two co-origin Gaussian functions}\label{SumCo-originGauss}
\begin{figure}[!h]
\centering
\includegraphics[width = 5cm]{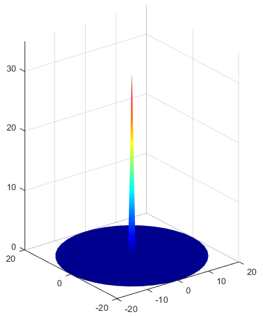}
\caption{Sum of two co-origin Gaussian Functions in real space (conceptual plot).}
\label{Fig:2D_Gaussian_Origin}
\end{figure}

An experiment involving the reconstruction of a sum of two two-dimensional Gaussian functions is performed in this section. The original function in real space is displayed in Figure \ref{Fig:2D_Gaussian_Origin} and has the following expression,
\begin{equation}
f(x,y)= \frac{1}{2\pi\sigma_1\sigma_2} \exp \left( -\frac{1}{2}\left(\frac{x^2}{\sigma_1^2} + \frac{y^2}{\sigma_2^2}\right)\right) + \frac{1}{2\pi\sigma_3\sigma_4} \exp \left( -\frac{1}{2}\left(\frac{x^2}{\sigma_3^2} + \frac{y^2}{\sigma_4^2} \right)\right)
\end{equation}

\noindent where $\sigma_1 = 0.1$, $\sigma_2 = 0.3$, $\sigma_3 = 0.05$ and $\sigma_4 = 0.25$.

\FloatBarrier

\begin{figure}[!h]
\centering
\includegraphics[width = 8cm]{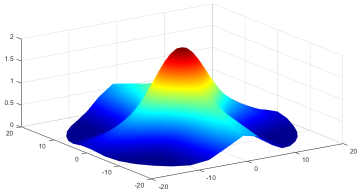}
\caption{Fourier transform of the sum of two co-origin Gaussian Functions in Fourier space (conceptual plot).}
\label{Fig:2D_Gaussian_Origin_FT}
\end{figure}

The Fourier transform of this function is the sum of another two Gaussian functions, as shown in Fig.~\ref{Fig:2D_Gaussian_Origin_FT}. The analytical expression for this Fourier transform would then be,

\begin{equation}
F(\omega_1,\omega_2) = \exp \left(\frac{-\sigma_1^2 \omega_1^2 -\sigma_2^2 \omega_2^2}{2}\right) +
\exp \left(\frac{-\sigma_3^2 \omega_1^2 -\sigma_4^2 \omega_2^2}{2}\right).
\end{equation}

\FloatBarrier

\begin{figure}[!h]
\centering
\includegraphics[width = 12cm]{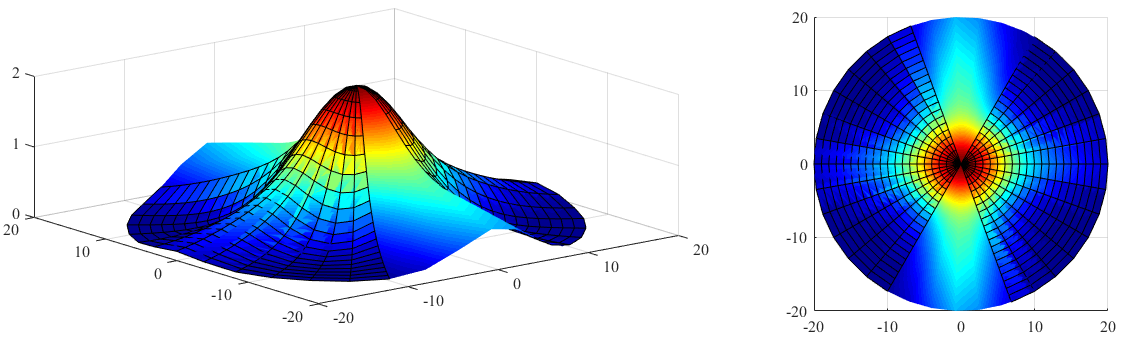}
\caption{Sampling of EM data}
\label{Fig:2D_Gaussian_sampling}
\end{figure}

A comparison of the Fourier transform of the original density with that obtained by SAXS-EM fusion and EM alone with the projection angle gap $[60^{\circ}, 110^{\circ}]$ is shown in Fig.~\ref{Fig:2D_Gaussian_sampling}. The known EM data is evenly distributed in the meshed range, while the unmeshed region indicates the gap in the data.

\FloatBarrier

\begin{figure}[!h]
\centering
\includegraphics[width = 15cm]{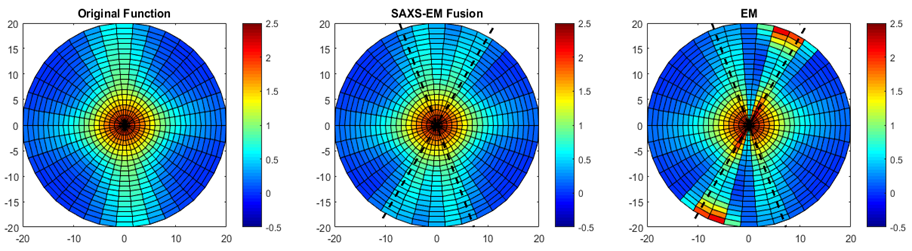}
\caption{Comparison of the Fourier transform of original density with that obtained by SAXS-EM fusion and EM alone}
\label{Fig:2D_Gaussian_Origin_reconstruction}
\end{figure}

Figure~\ref{Fig:2D_Gaussian_Origin_reconstruction} shows the result with the projection angle gap [60, 110]. As can be seen from the figure, SAXS-EM fusion gives results close to the original function while EM alone cannot reconstruct the values well. This result therefore shows a success of the synergism between cryo-EM and SAXS in the 2D case. More tests were run with different original functions, input angles, regions and widths of gaps, showing consistent results.

\FloatBarrier


\subsection{2D smiley face}\label{Smile2D}
\subsubsection{Numerical experiment procedure}

\begin{figure}[!h]
  \centering
  \includegraphics[width=15cm]{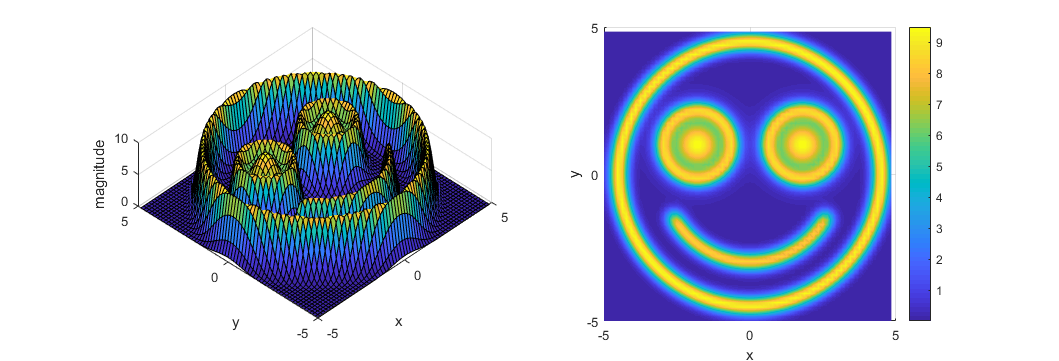}
  \caption{Smiley Face constructed by sum of 2D Gaussian distributions}\label{Fig:2D_smile_real}
\end{figure}

A two-dimensional smiley face was constructed using (\ref{NGaussian_Combine}) and substituting $N = 2$ in the definition of the Gaussian in (\ref{NGaussian}). Here, the mean vector, $\bm{\mu}_k$ would represent the two-dimensional coordinates of a point on a discretized smiley face, with each coordinate in the smiley face mapping to a $\bm{\mu}_k$ and a corresponding Gaussian, $f(\bm{x};\bm{\mu}_k,\Sigma_k)$. The weighting coefficient $A_k$ was adjusted for each Gaussian to modify the amplitude of each function and a smiley face was created as shown in Fig.~\ref{Fig:2D_smile_real}. The analytical Fourier transform, $F(\bm{\omega})$ was evaluated for each Gaussian using (\ref{NGaussian_Fourier}) with $N = 2$ and summed using (\ref{NGaussian_Fourier_Combine}). 

\FloatBarrier

\begin{figure}[!h]
\centering
\begin{minipage}[t]{0.48\textwidth}
\centering
\includegraphics[width=6cm]{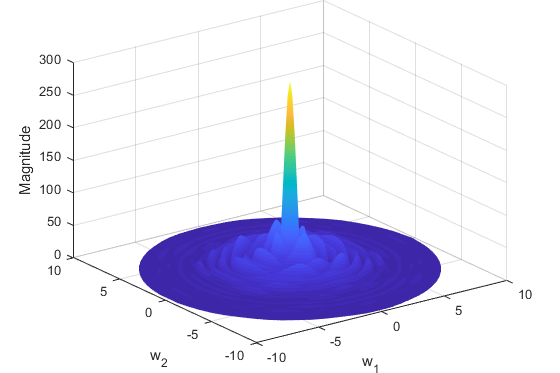}
\caption{Geometric situation of Smiley Face after Fourier transform}
\label{Fig:2D_smile_fourier}
\end{minipage}
\begin{minipage}[t]{0.48\textwidth}
\centering
\includegraphics[width=6cm]{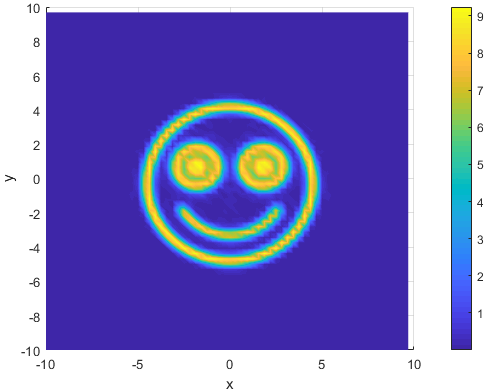}
\caption{Reconstruction of 2D smiley face with no projection gap}
\label{Fig:smile_ground_truth}
\end{minipage}
\end{figure}

Since the SAXS experiment provides the square average of the function $F(\bm\omega)$ along circles with different radii, we let $\bm{\omega}=[r\cos\theta,r\sin\theta]$; the Fourier transform plotted in polar coordinates is shown in Fig.~\ref{Fig:2D_smile_fourier}. Then, the SAXS data can be obtained by integrating the square of the amplitude of the analytical Fourier transform along each circle with radius $r$. Similar to \eqref{eq_12}, the SAXS information is,
\begin{equation}
I(r) = \int_{0}^{2\pi} |F(r,\theta)|^2 \mathrm{d} \theta \,.
\label{eq:smile2DSAXSintegral}
\end{equation}

\FloatBarrier

In our numerical experiment, the Fourier transform of the original smiley face is discretized into 65 circles whose radii are evenly spaced. Along each circle, 65 points are distributed evenly. Following this, a gap is introduced in the projection directions in order to simulate the EM experiment. All numerical simulations were performed with a MATLAB R2018b, i5-4690 CPU.

Both the EM reconstruction and SAXS-EM fusion reconstruction were completed along each circle where a quadratic (\ref{eq:quadratic_2Dsimple}) was solved. Equation ~\eqref{eq:Quadroots} gives the analytical expression of the root. But instead of selecting the true root by selectively blocking data obtained at various angles and choosing the fairly stable one, we obtain an approximate value of the root calculated by \eqref{eq:DirectSumApprox2D} as a reference to to select root. The one closer to this approximated value is chosen as the right root. 
 On the other hand, if the roots are complex, we only take the real part of the root. During the numerical experiment, when the radius in Fourier space gets larger, the root tends to transform from real to complex, introducing relatively large errors when compared with the analytical solution for the true roots. But in a real EM experiment, we have no way to do such a comparison with an analytical result because only discrete data from EM is provided. Therefore, the root solved from the quadratic equation that is closest to the approximated selecting root is the one that we use to fuse the SAXS and EM information.

\subsubsection{Results and discussion}
\paragraph{$30^{\circ}$ Projection Angle Gap}~
\newline

\begin{figure}[H]
  \centering
  \includegraphics[width=12cm]{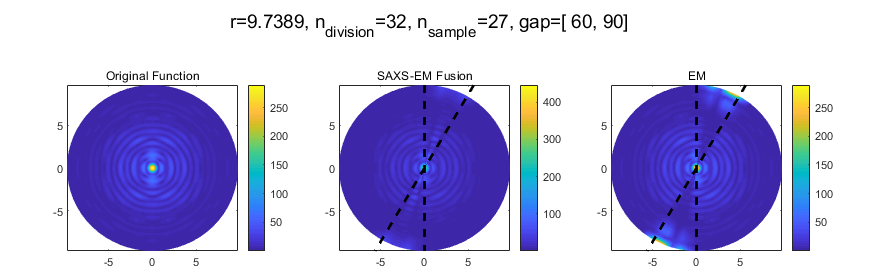}
  \caption{Comparison of Fourier transform of original density with that obtained by SAXS-EM
fusion and EM alone}\label{fig:smile2dcompare3}
\end{figure}

Figure \ref{Fig:smile_ground_truth} provides the ground truth of the smiley face reconstruction using complete EM data alone (\textit{i.\,e.} there are no missing wedges in the EM data). Figure \ref{fig:smile2dcompare3} shows the result with the projection angle gap $[60^{\circ}, 90^{\circ}]$. As can be seen from the figure, SAXS-EM fusion restores most information in the projection gap that EM alone cannot. EM data alone also gives a false restoration at the edge of the region. These high frequency artefacts in Fourier space correspond to sharp discontinuities or changes in the reconstructed smiley face in real space. The presence of such artefacts can also lower the resolution of the final reconstructed smiley face.

\FloatBarrier

With the results provided by the two methods above, a real smiley face can be reconstructed by interpolating from the discrete data in polar coordinates to Cartesian coordinates to set-up the data for an inverse FFT. To improve the accuracy of coordinate transformation, oversampling is used here to increase the density of radius and angle distribution. Based on the intensive data, two-dimensional cubic spline interpolation is used to evaluate the value at each point given in the Cartesian coordinate. The output domain of the polar to Cartesian transformation and interpolation is a square whose side length equals to the diameter of the biggest circle that can be drawn in the polar coordinates; outside this circle, the Fourier transform values are set to zero, consistent with the analytical condition.

\begin{figure}[!h]
  \centering
  \includegraphics[width=12cm]{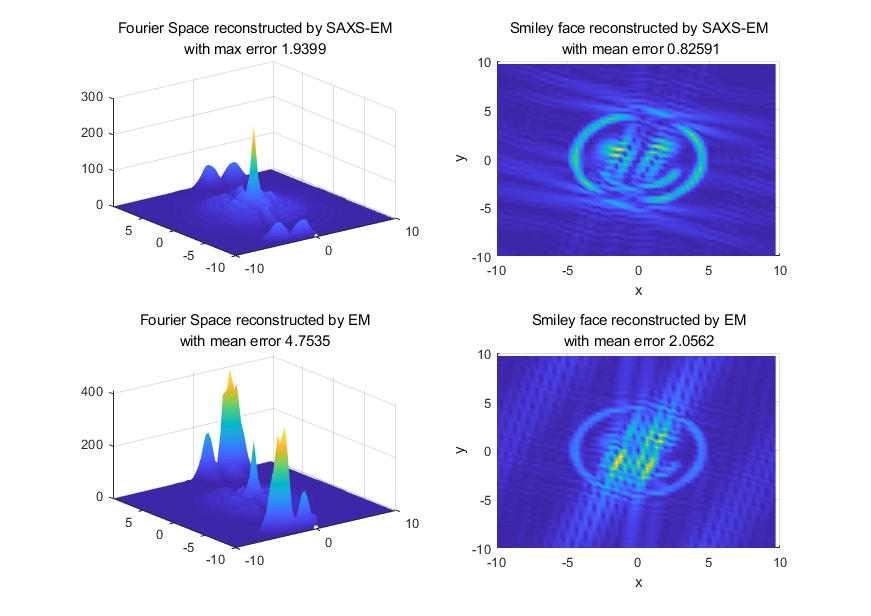}
  \caption{Comparison of smiley face reconstruction obtained by SAXS-EM
fusion and EM alone for $[60^{\circ},90^{\circ}]$ missing wedge }\label{fig:smile2dreconstruct}
\end{figure}

Figure \ref{fig:smile2dreconstruct} gives the results and comparison of reconstruction of the smiley face obtained by SAXS-EM fusion and EM alone. In Fourier space, the EM reconstruction adds two more peaks in addition to the central peak, resulting in large errors when filling the missing wedge. The mean error in magnitude compared of the SAXS-EM result compared with the analytical smiley face Gaussian distribution is 0.82591 while EM alone without the SAXS fusion gives an error of 2.0562 when compared with the analytical smiley face result. This shows a success of the synergism between cryo-EM and SAXS in 2D case when reconstructing to real space.

\begin{figure}[!h]
  \centering
  \includegraphics[width=12cm]{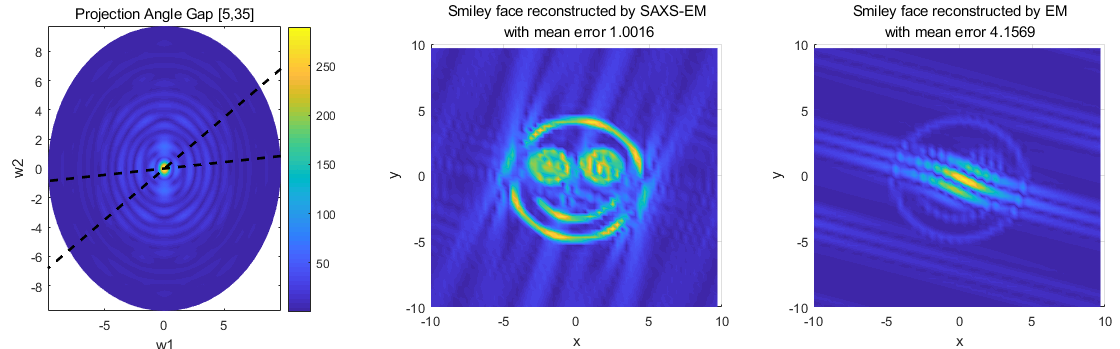}
  \caption{Comparison of smiley face reconstruction obtained by SAXS-EM
fusion and EM alone $[5^{\circ},35^{\circ}]$ missing wedge }\label{fig:smile2d_5_30}
\end{figure}

\begin{figure}[!h]
  \centering
  \includegraphics[width=12cm]{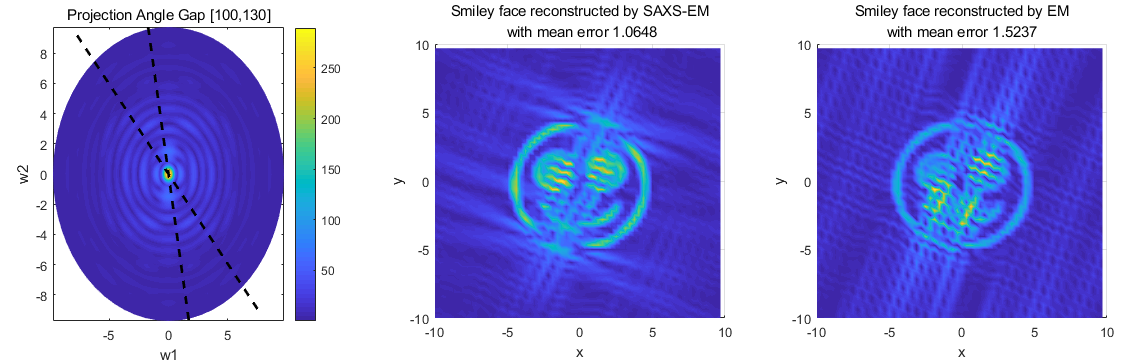}
  \caption{Comparison of smiley face reconstruction obtained by SAXS-EM
fusion and EM alone under $[100^{\circ},130^{\circ}]$ missing wedge }\label{fig:smile2d_100_30}
\end{figure}

Figure \ref{fig:smile2d_5_30} depicts the results and comparison of reconstruction with a $30^{\circ}$ projection angle gap when the start angle is $5^{\circ}$. The result shows a higher error in SAXS-EM fusion reconstruction in the case when information in $[5^{\circ}, 35^{\circ}]$ is missing as opposed to that when $[60^{\circ}, 90^{\circ}]$ is missing; nevertheless, the SAXS-EM fused result outperforms the EM-only reconstruction. Figure \ref{fig:smile2d_100_30} shows the results and comparison of a reconstruction with a $30^{\circ}$ projection angle gap when the start angle is $100^{\circ}$. In such a condition, the SAXS-EM fusion does not show any distinctive advantage and both EM alone and SAXS-EM reconstruct a relatively good smiley face. In conclusion, the reconstruction effect is related to the information contained in the projection gap. Only when a certain amount of key information is contained in the gap will the advantage of SAXS-EM fusion be more pronounced.

\FloatBarrier

\FloatBarrier

\FloatBarrier

\paragraph{Two Projection Angle Gaps}~
\newline

\begin{figure}[!h]
  \centering
  \includegraphics[width=12cm]{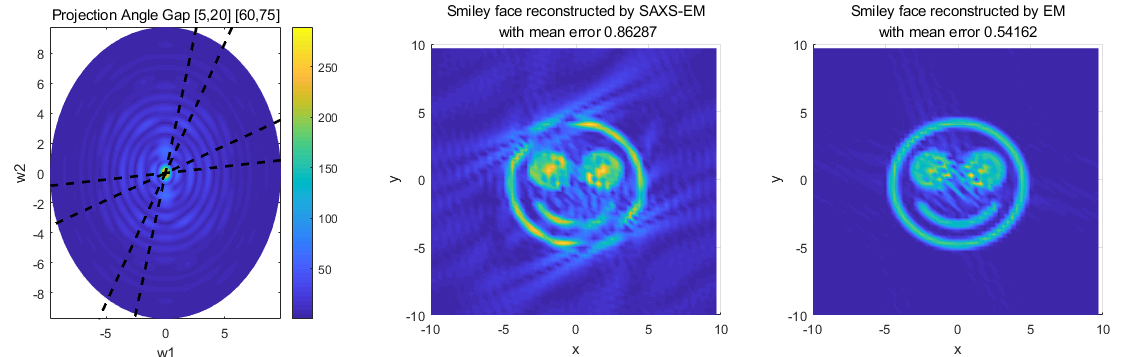}
  \caption{Comparison of smiley face reconstruction obtained by SAXS-EM fusion and EM alone under two gaps of $15^{\circ}$ each.}\label{fig:smile2d_two_gaps}
\end{figure}

\begin{figure}[!h]
  \centering
  \includegraphics[width=12cm]{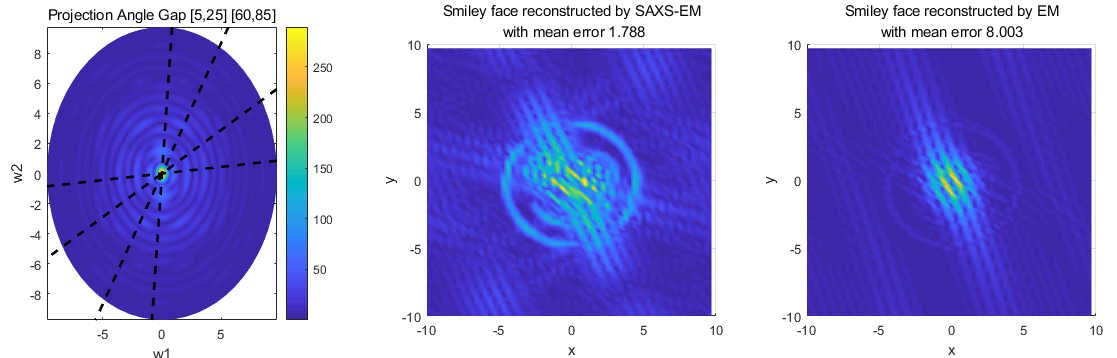}
  \caption{Comparison of smiley face reconstruction obtained by SAXS-EM
fusion and EM alone under two gaps of $25^{\circ}$ each.}\label{fig:smile2d_two_gaps_25}
\end{figure}

Up to this point, we have investigated the case of one gap in the projection angle. In this subsection, two projection angle gaps of $15^{\circ}$ and $25^{\circ}$ each are considered. Figure \ref{fig:smile2d_two_gaps} shows the result when the gaps are $[5^{\circ},20^{\circ}]$ and $[60^{\circ},75^{\circ}]$. The EM experiment helps to reconstruct a smoother smiley face while SAXS-EM fusion provides additional artefacts that result in a higher mean error. But judging from the color, the SAXS-EM fusion result is closer to the ground truth in the outline and the eyes of the face. Since $15^{\circ}$ is so small that each gap does not contain much information, we increased the missing wedge and tested $[5^{\circ},30^{\circ}]$ and $[60^{\circ},85^{\circ}]$. Figure \ref{fig:smile2d_two_gaps_25} shows the reconstruction result of two wedges when each angle gap is $25^{\circ}$. By using SAXS-EM fusion, most of the smiley face is reconstructed except for some diagonal streaks in the centre, which are also present in the EM results. Nevertheless, SAXS-EM fusion has a considerably lower error. This illustrates again that the difference between a SAXS-EM fusion reconstruction and an EM-alone reconstruction is greater if the missing wedges contain important information about the object.

\FloatBarrier

\FloatBarrier

\newpage
\paragraph{Larger Gaps and Fewer Sampling points}~
\newline

\begin{figure}[!h]
  \centering
  \includegraphics[width=12cm]{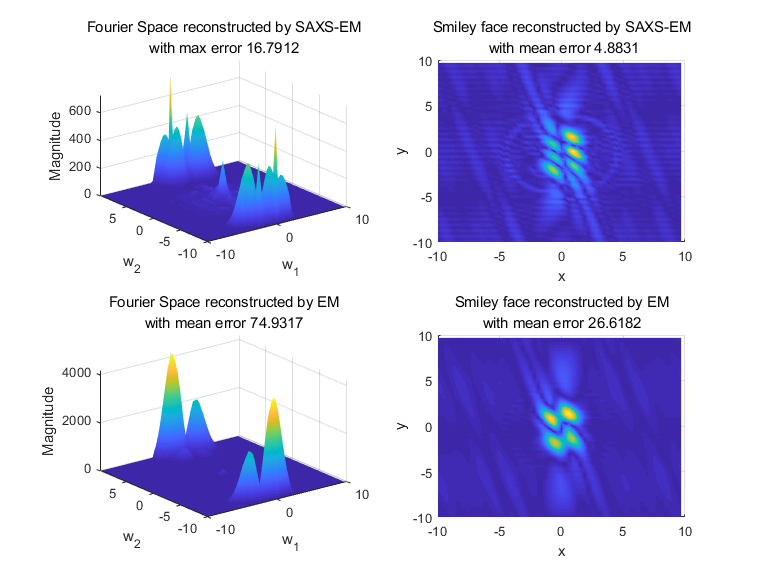}
  \caption{Comparison of smiley face reconstruction obtained by SAXS-EM
fusion and EM alone under $50^{\circ}$ missing wedge}\label{fig:smile2d6450}
\end{figure}

In the cases of a larger projection angle gap such as $[60^{\circ},110^{\circ}]$, Fig.~\ref{fig:smile2d6450} gives the results where both SAXS-EM and EM-alone reconstructions fail to reconstruct a smiley face. In Fourier space, we find that the largest error occurs when the points in the gap are far from the centre. However, the reconstructed smiley face using SAXS-EM information fusion still has some advantages due to the display of the face outline and the lower mean error compared with that given by using EM data alone. In other cases, when the projection angle gap gets smaller (\textit{i.\,e.} $10^{\circ}$), the two methods can both reconstruct the 2D smiley face successfully with low error.

Another remark is that the reconstruction effect is also related to grid size in real space. When we sample 33 different radii and 33 discrete points evenly spaced along each circle, SAXS-EM fusion fails to show its superiority in the smiley face reconstruction. This is partly because of an ill-conditioning in the matrix, $M^{-1} = (FF^{H})$, leads to numerical errors in the computation of $M$ in (\ref{eq:quadratic_2Dsimple}) at certain angle gaps and starting angles. Moreover, since the root selection procedure is not entirely foolproof, it is possible that for some radii a ``wrong" root is selected in a coarse grid ---where the sampling is not fine enough to ensure that both roots have converged---allowing the possibility of the SAXS reconstruction performing worse than the EM reconstruction.

\FloatBarrier

\begin{figure}[!h]
  \centering
  \includegraphics[width=12cm]{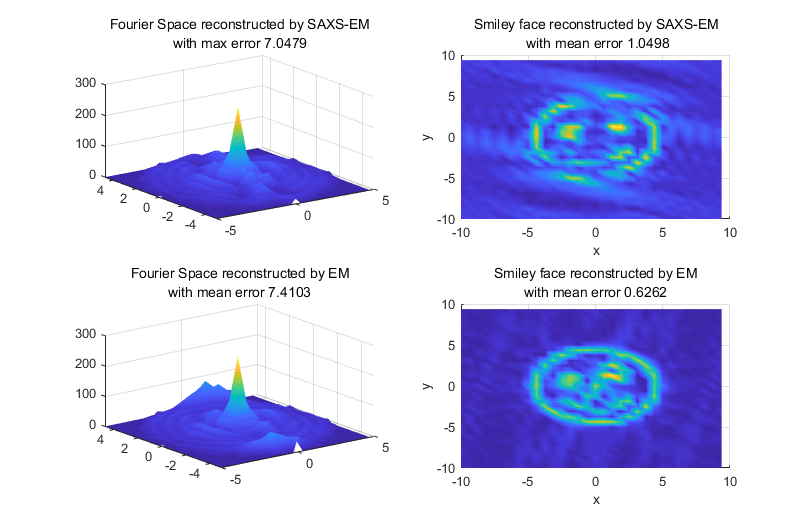}
  \caption{Comparison of smiley face reconstruction obtained by SAXS-EM
fusion and EM alone under $50^{\circ}$ missing wedge}\label{fig:smile2d3250}
\end{figure}
\FloatBarrier

\subsection{3D smiley face}\label{3DSmile}
\subsubsection{Numerical experiment procedure}

\begin{figure}[!h]
\centering
\includegraphics[width = 5cm]{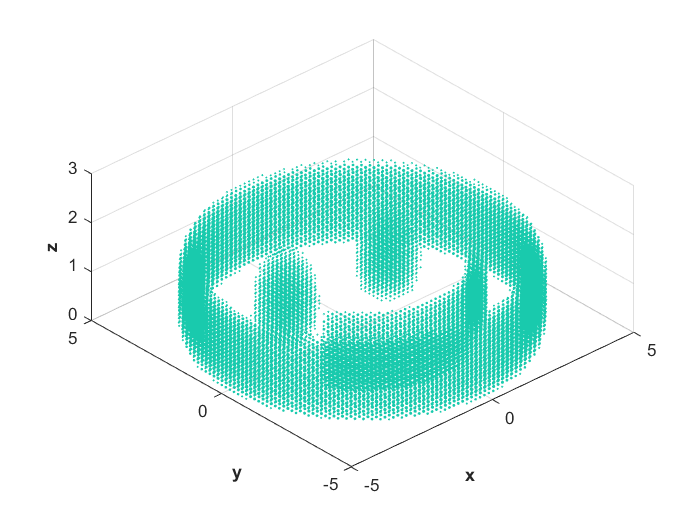}
\caption{3D smiley face}
\label{Fig:3D_smileface}
\end{figure}

A three-dimensional smiley face was constructed using (\ref{NGaussian_Combine}) and substituting $N = 3$ in the definition of the Gaussian in (\ref{NGaussian}). Here, the $\boldsymbol{\mu}$ of each Gaussian distribution is similar to that in the two-dimensional case in Sec.~\ref{Smile2D}. The only difference between the two-dimensional and three-dimensional mean vector is in the component of $\bm{\mu}$ perpendicular to the plane containing the two-dimensional smiley face explored in Sec.~\ref{Smile2D}. In this example, we construct a three-dimensional smiley face such that $\bm{\mu}=(\mu_x, \mu_y, \mu_z)$ and $(\mu_x, \mu_y)$ is the mean vector corresponding to a given Gaussian on the two-dimensional smiley face in Sec.~\ref{Smile2D}. Then, $\mu_z$ is defined as 10 values divided evenly from 0 to 1.4. In effect, the three-dimensional smiley face is constructed by extruding the two-dimensional smiley face and discretizing it into spherical Gaussians; the result would be a cylindrical object as shown in Fig.~\ref{Fig:3D_smileface}. The Fourier transform of the three-dimensional smiley face is then calculated by (\ref{NGaussian_Fourier}) with $N = 3$ and summed using (\ref{NGaussian_Fourier_Combine}).

\FloatBarrier

\begin{figure}[!h]
\centering
\includegraphics[width = 10cm]{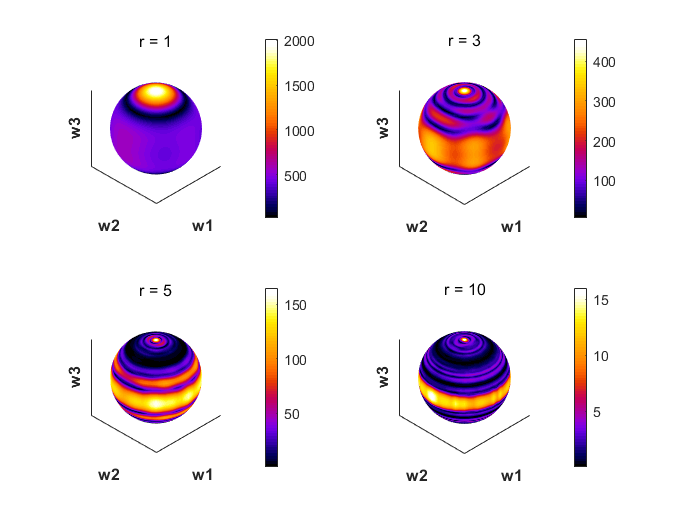}
\caption{Fourier transform of the smiley face ($r = 1, 3, 5, 10$)}
\label{Fig:3D_smile_face_fourier_space_each_r}
\end{figure}

The Fourier transform of the smiley face in spherical coordinate for several values of radius can be seen in Fig.~\ref{Fig:3D_smile_face_fourier_space_each_r}.

\FloatBarrier

\begin{figure}[!h]
\centering
\includegraphics[width = 5cm]{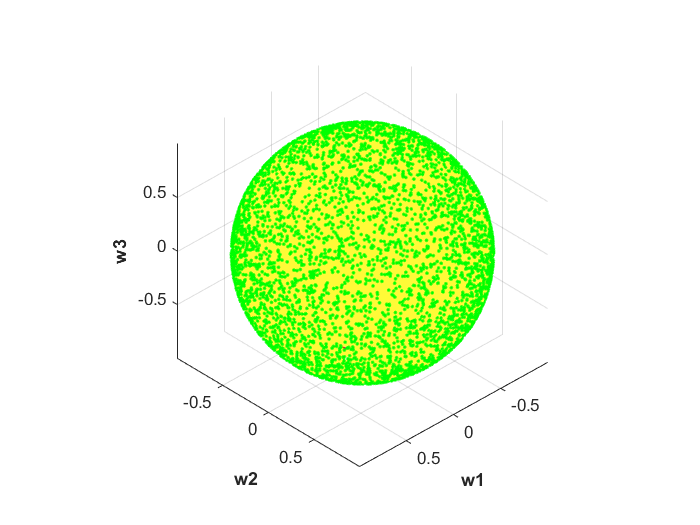}
\caption{Sampled points in Fourier space}
\label{Fig:3D_smile_face_sample_points_fourier_space_singlercolor_sphere}
\end{figure}

Since it is possible to deduce the Fourier transform of the 3D smiley face analytically, we can generate multiple great circles and obtain the value of the Fourier transform on sampled points on each great circle using the analytical function. This sampling was done at discrete points and great circles to simulate the incompleteness of the data generated by the EM experiment. Figure~\ref{Fig:3D_smile_face_sample_points_fourier_space_singlercolor_sphere} shows all the sampled points from all great circles considered in Fourier space. At the moment, the number of sampled points for each sphere with different radii in Fourier space is the same but this need not always be the case, and the case where the number of sampled points is a non-constant function of sphere radius will be discussed later.

\FloatBarrier

\begin{figure}[!h]
\centering
\begin{minipage}[t]{0.48\textwidth}
\centering
\includegraphics[width=7cm]{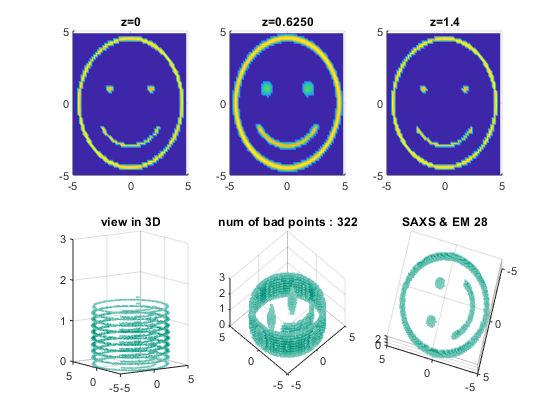}
\caption{Result of SAXS-EM reconstruction}
\label{Fig:3D_smile_face_SAXS_EM_reconstruction_without_remove}
\end{minipage}
\begin{minipage}[t]{0.48\textwidth}
\centering
\includegraphics[width=7cm]{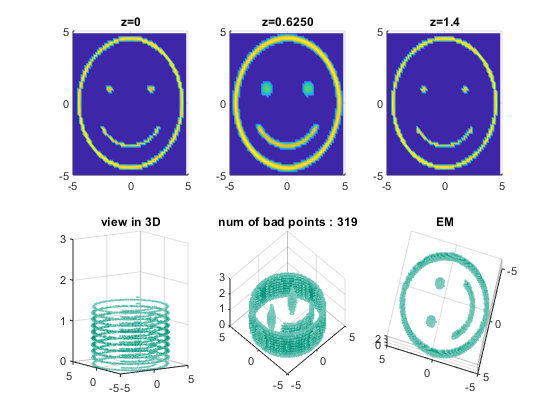}
\caption{Result of EM reconstruction}
\label{Fig:3D_smile_face_EM_reconstruction_without_remove}
\end{minipage}
\end{figure}

Now, we proceed to discuss the results obtained in Fourier space by SAXS-EM fusion and an EM-alone experiment. With that, a real smiley face can be reconstructed by interpolating from discrete data in spherical coordinate to Cartesian coordinate that is prepared for inverse FFT. Based on the data in spherical coordinates, a three-dimensional cubic spline interpolation is used to evaluate the value at each point given in Cartesian coordinates. Like the 2D case (\ref{Smile2D}), the whole region of Cartesian coordinate is a cube whose side length equals the diameter of the biggest circle given in spherical coordinates whereas outside the spherical range the value of the Fourier transform is set to zero, consistent with the analytical condition. After that, we take the inverse Fourier transform to real space. The reconstruction obtained by the two methods is showed in Fig.~\ref{Fig:3D_smile_face_SAXS_EM_reconstruction_without_remove} and Fig.~\ref{Fig:3D_smile_face_EM_reconstruction_without_remove}. In Fig.~\ref{Fig:3D_smile_face_SAXS_EM_reconstruction_without_remove} and Fig.~\ref{Fig:3D_smile_face_EM_reconstruction_without_remove}, the first row shows three slices of the reconstruction to make a better comparison and the second row is the 3D model in a different view. To show a better comparison, we define ``bad points" as the points at which the error between the original and the reconstructed function is greater than 20\% of the original value.

We can see that the two reconstructions are all good since there are many sampled points in Fourier space and the simulation covers almost all the information. However, we may lose some information in a real experiment. 

\subsubsection{Results and discussion}
\paragraph{Random Great Circles with Constant Sampling Number}~
\newline

\begin{figure}[!h]
\centering
\begin{minipage}[t]{0.48\textwidth}
\centering
\includegraphics[width=6cm]{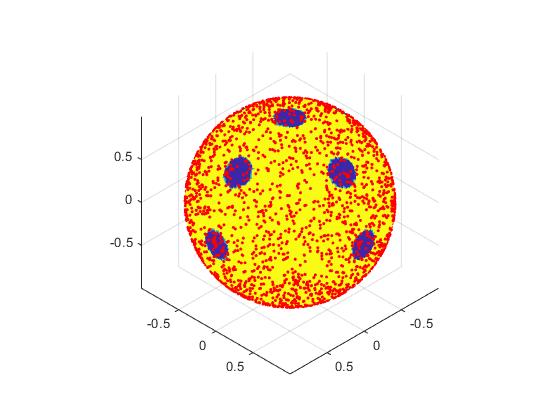}
\caption{Direction of great circles removed}
\label{Fig:3D_smile_face_remove_direction}
\end{minipage}
\begin{minipage}[t]{0.48\textwidth}
\centering
\includegraphics[width=6cm]{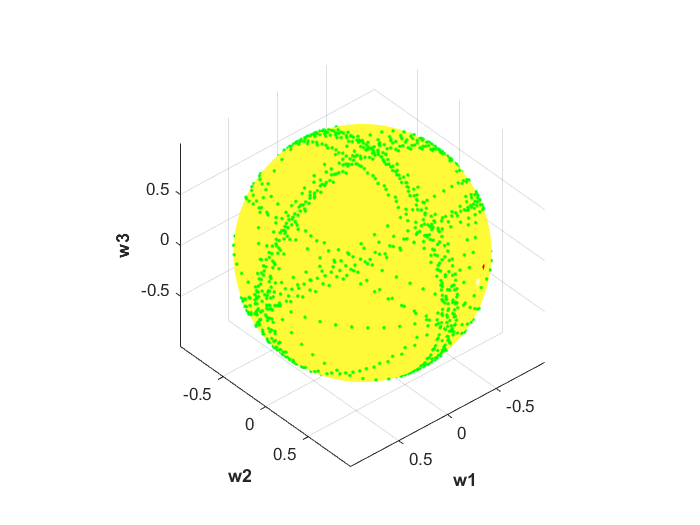}
\caption{Sample points after exclusion}
\label{Fig:3D_smile_face_sample_direction_remove_singlecolor_space}
\end{minipage}
\end{figure}

To simulate what happens in the real experiment, we also set some cones in the sphere in Fourier space and exclude all the great circles whose directions are outside these cones. As shown in Fig.~\ref{Fig:3D_smile_face_remove_direction}, the red points represent the direction perpendicular to the plane containing the great circles and the blue cones represent the collection of those directions whose corresponding great circles are sampled in the EM data. Those red points which are not in the blue cones are excluded. The center axes of these blue cones are generated by several different sets of $\theta \text { and } \varphi$, where

\begin{equation*}
\begin{array}{l}{\varphi \in\left[\begin{matrix}
   0 & \frac{\pi}{2} & \pi 
  \end{matrix} \right]} \\ {\theta \in\left[\begin{matrix}
   0 & \frac{\pi}{4} & \frac{\pi}{2} &\frac{3 \pi}{4}& \pi 
  \end{matrix} \right].}\end{array}
\end{equation*}

The angle around each center axes is $9^{\circ}$.

\FloatBarrier

\begin{figure}[!h]
\centering
\begin{minipage}[t]{0.48\textwidth}
\centering
\includegraphics[width=7cm]{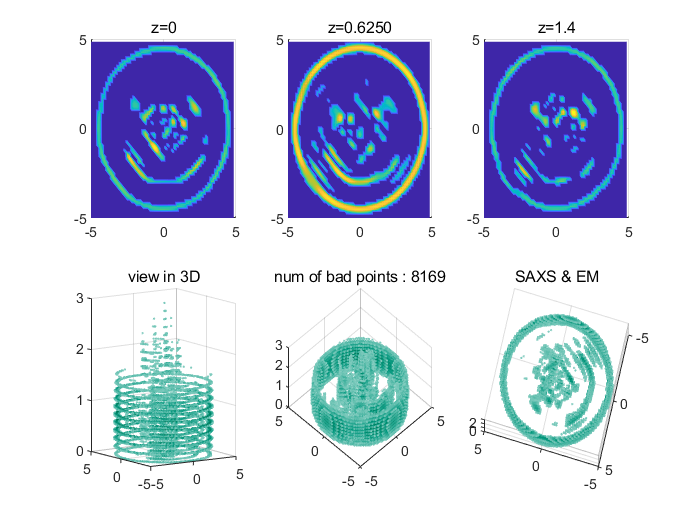}
\caption{SAXS-EM reconstruction by random great circle and constant sampling points}
\label{Fig:3D_smile_face_direction_remove_SAXS_EM}
\end{minipage}
\begin{minipage}[t]{0.48\textwidth}
\centering
\includegraphics[width=7cm]{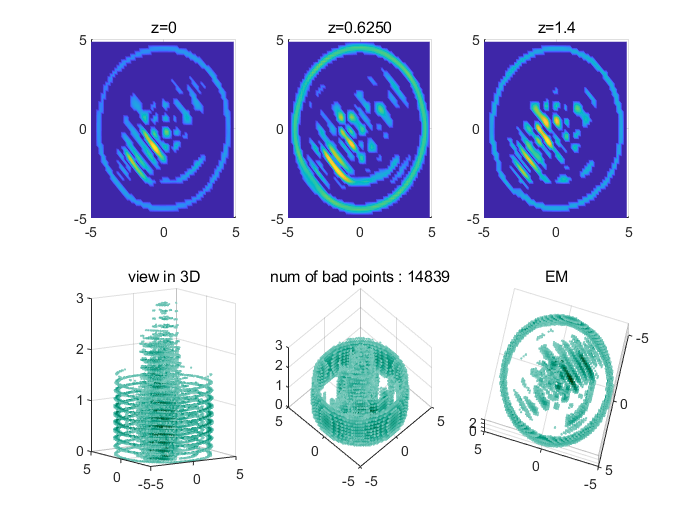}
\caption{EM reconstruction by random great circle and constant sampling points} \label{Fig:3D_smile_face_direction_remove_EM}
\end{minipage}
\end{figure}

The reconstruction result by two methods are showed in Figure~\ref{Fig:3D_smile_face_direction_remove_EM} and Figure~\ref{Fig:3D_smile_face_direction_remove_SAXS_EM}.

\FloatBarrier

The reconstruction by SAXS-EM fusion is more like a smiley face than that obtained by EM alone especially in $z=0$ plane. Though we may not see an obvious difference between two results, the number of bad points can show the advantage of SAXS-EM fusion.

\paragraph{Uniformly Distributed Great Circles with Constant Sampling Number}~
\newline

\begin{figure}[!h]
\centering
\begin{minipage}[t]{0.48\textwidth}
\centering
\includegraphics[width=6cm]{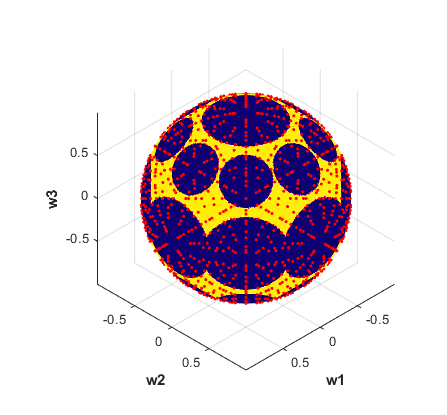}
\caption{Normal direction of great circles removed}
\label{Fig:3D_smile_face_normal_vector_even}
\end{minipage}
\begin{minipage}[t]{0.48\textwidth}
\centering
\includegraphics[width=6cm]{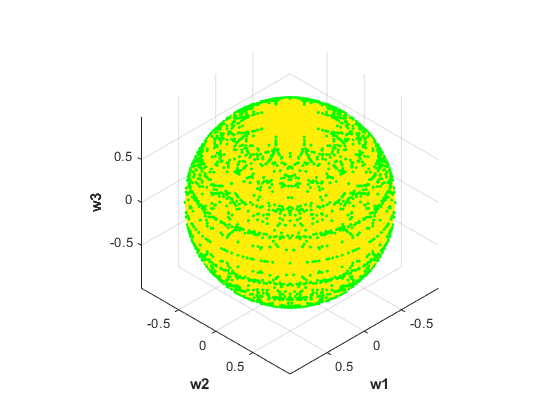}
\caption{Sample points after exclusion}
\label{Fig:3D_smile_face_sample_points_even_vector}
\end{minipage}
\end{figure}

In the simulation above, since the normal direction of great circles are generated randomly, the normal vectors will tend aggregate in some areas and be sparsely distributed in other regions. Hence, we generate the great circles evenly on the sphere, as shown in Fig.~\ref{Fig:3D_smile_face_normal_vector_even}, and remove the information obtained by great circles whose normal vectors, which are the red points, are in the blue cones.

\FloatBarrier

\begin{figure}[!h]
\centering
\begin{minipage}[t]{0.48\textwidth}
\centering
\includegraphics[width=7cm]{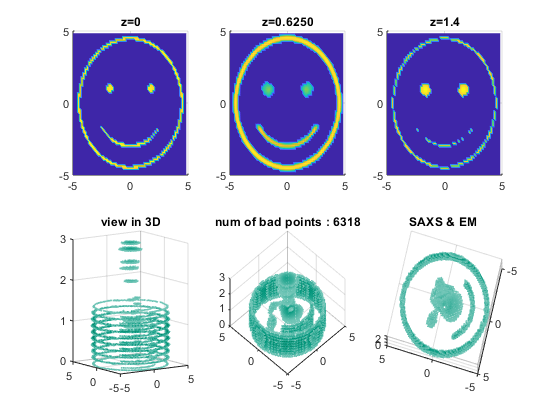}
\caption{SAXS-EM reconstruction by uniform great circles and constantly sampling}
\label{Fig:3D_smile_face_even_great_circle_SAXS_EM_reconstruction}
\end{minipage}
\begin{minipage}[t]{0.48\textwidth}
\centering
\includegraphics[width=7cm]{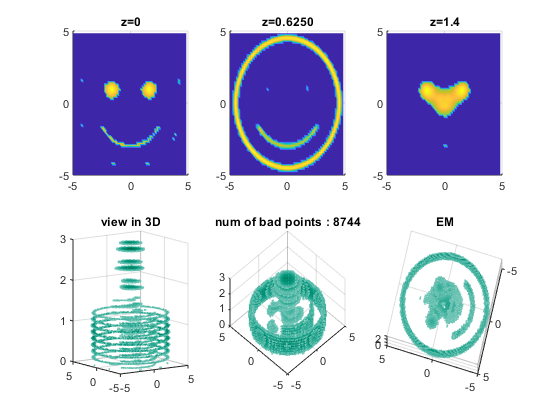}
\caption{EM reconstruction by uniform great circles and constantly sampling}
\label{Fig:3D_smile_face_even_great_circle_EM_reconstruction}
\end{minipage}
\end{figure}

With the same process mentioned above, we could get the reconstruction of smiley face in real space. The  result obtained by SAXS-EM and EM alone can be seen in Fig.~\ref{Fig:3D_smile_face_even_great_circle_SAXS_EM_reconstruction} and Fig.~\ref{Fig:3D_smile_face_even_great_circle_EM_reconstruction}. 
\FloatBarrier

We can see that the outline of the smiley face is missed in plane $z=0$ and $z=1.4$ and the helix above the smiley face, which is an artefact, is denser when reconstructing using only EM.

\paragraph{Uniformly Distributed Great Circles with Radius Dependent Sampling Number}~
\newline

In the process of reconstruction above, the sampled points for each great circle in Fourier space is constant for each sphere with different radii. Since the perimeter of a great circle is linear with respect to the radius, if we take sample points like that, we may miss much information in sphere with a larger radius and get too much information in sphere with smaller radius, which is not true in the real experiment.

\begin{figure}[!h]
\centering
\includegraphics[width = 13cm]{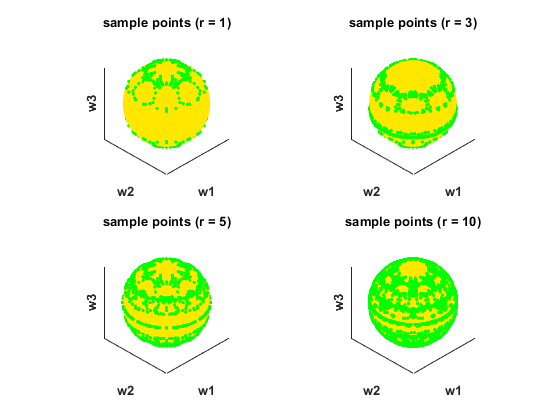}
\caption{Sampled points in sphere with different radius}
\label{Fig:3D_smile_face_sample_points_different_r}
\end{figure}

Hence, we modify the number of sample points in each great circle to be linear to the radius of the sphere. We take the same great circles and missing wedges as showed in Fig.~\ref{Fig:3D_smile_face_normal_vector_even}. In this case, the number of the sampled points in each great circles is linear to the radius, $ num = \lceil kr \rceil$. We set $k = 1.5$ in our first simulation and the sampled points in sphere with different radius are showed in Figure~\ref{Fig:3D_smile_face_sample_points_different_r}.

\FloatBarrier

\begin{figure}[!h]
\centering
\begin{minipage}[t]{0.48\textwidth}
\centering
\includegraphics[width=7cm]{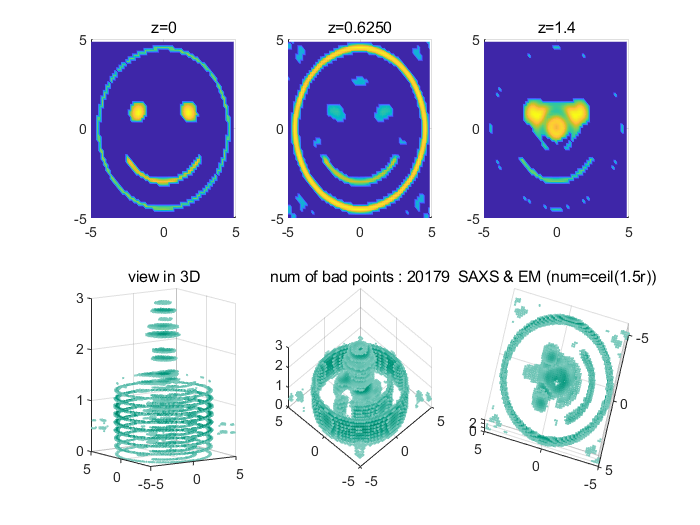}
\caption{SAXS-EM reconstruction (k=1.5)}
\label{Fig:3D_smile_face_SAXS_EM1_5}
\end{minipage}
\begin{minipage}[t]{0.48\textwidth}
\centering
\includegraphics[width=7cm]{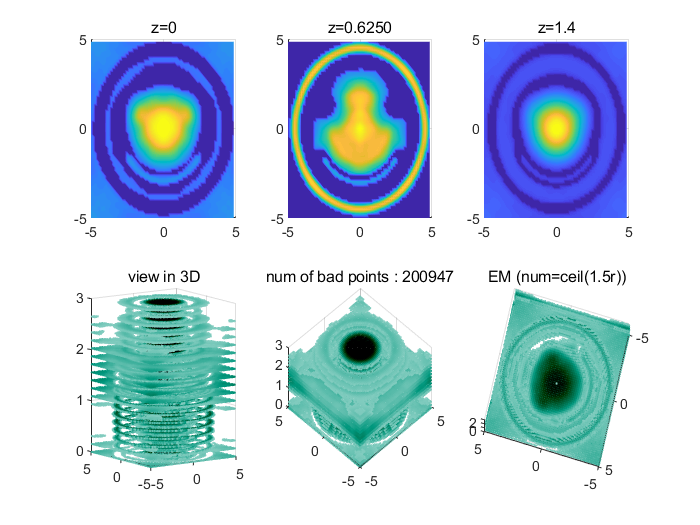}
\caption{EM reconstruction (k=1.5)}
\label{Fig:3D_smile_face_EM1_5}
\end{minipage}
\end{figure}

With same process, we can get the reconstruction of two methods and the results obtained by SAXS-EM and EM alone are in Fig.~\ref{Fig:3D_smile_face_SAXS_EM1_5} and  Fig.~\ref{Fig:3D_smile_face_EM1_5}.

We can see that the reconstruction obtained by SAXS-EM looks quite similar to the original one while we can hardly say that the reconstruction obtained by EM alone looks like a smiley face. Also, the number of bad points of the reconstruction by using EM alone is much greater than that by using SAXS-EM fusion. 

\begin{figure}[!h]
\centering
\begin{minipage}[t]{0.48\textwidth}
\centering
\includegraphics[width=7cm]{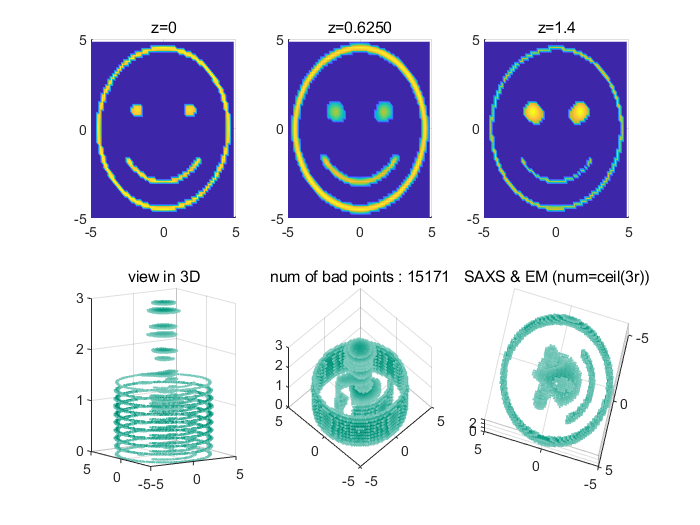}
\caption{SAXS-EM reconstruction  (k=3)}
\label{Fig:3D_smile_face_SAXS_EM_r3}
\end{minipage}
\begin{minipage}[t]{0.48\textwidth}
\centering
\includegraphics[width=7cm]{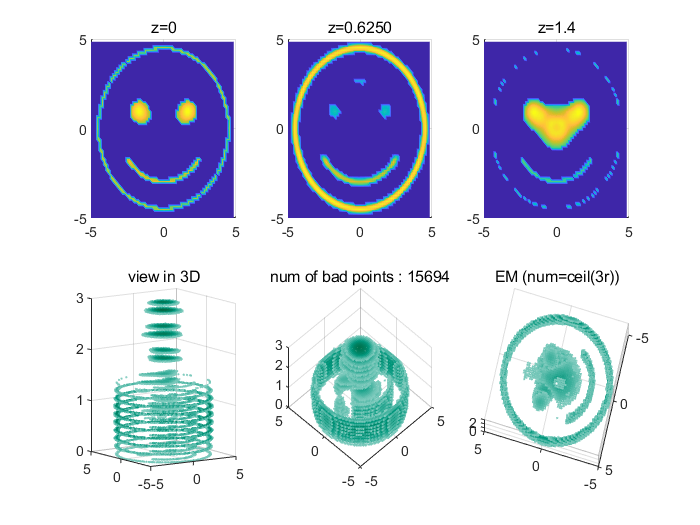}
\caption{EM reconstruction  (k=3)}
\label{Fig:3D_smile_face_EM_r3}
\end{minipage}
\end{figure}

At a higher value of $k$, when $k = 3$, there is a greater improvement in the results. In this case, for the SAXS-EM reconstruction, one can clearly see the smiley face. Using EM alone, we can also get a good reconstruction since we obtain enough information with a higher value of $k$. The results are shown in Fig.~\ref{Fig:3D_smile_face_SAXS_EM_r3} and Fig.~\ref{Fig:3D_smile_face_EM_r3}.

\FloatBarrier

\begin{figure}[!h]
\centering
\begin{minipage}[t]{0.48\textwidth}
\centering
\includegraphics[width=7cm]{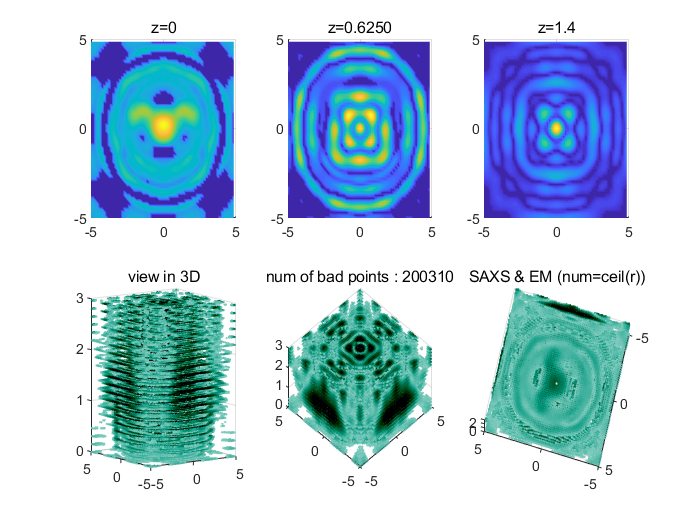}
\caption{SAXS-EM reconstruction  (k=1)}
\label{Fig:3D_smile_face_SAXS_EM_r1}
\end{minipage}
\begin{minipage}[t]{0.48\textwidth}
\centering
\includegraphics[width=7cm]{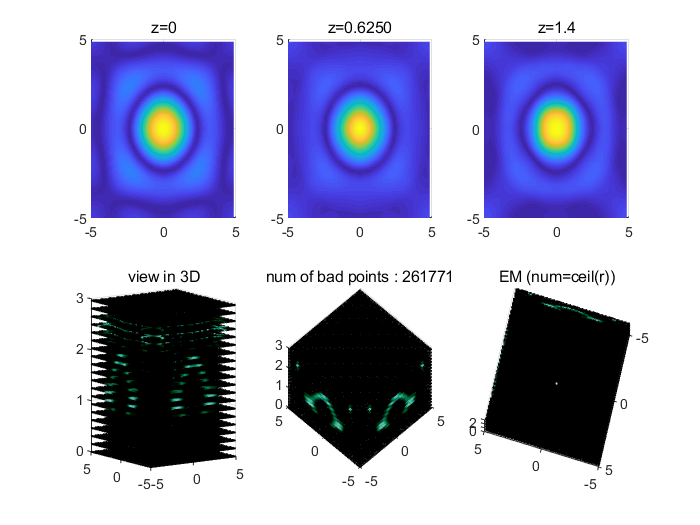}
\caption{EM reconstruction  (k=1)}
\label{Fig:3D_smile_face_EM_r1}
\end{minipage}
\end{figure}

A lower value of $k$, where $k = 1$, also been simulated. The result can be seen in Fig.~\ref{Fig:3D_smile_face_SAXS_EM_r1} and Fig.~\ref{Fig:3D_smile_face_EM_r1}. In this case, the EM reconstruction looks messy. Moreover, the SAXS-EM reconstruction also does not look like a smiley face. Nevertheless, we can see the outline of the smiley face in SAXS-EM reconstruction and the number of bad points is also much less in the SAXS-EM reconstruction.

\FloatBarrier

\subsection{Minion}

\begin{figure}[!h]
    \centering
    \includegraphics[width=8cm]{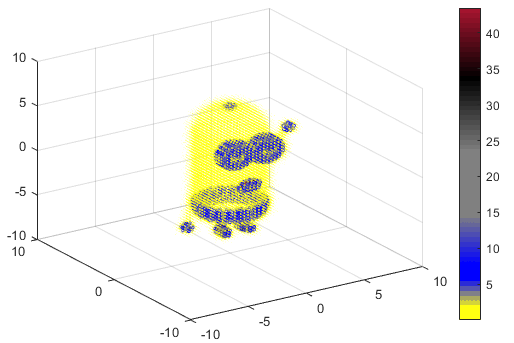}
    \caption{Minion created in real space}
    \label{fig:minions_real}
\end{figure}

Before we apply the technique to a real biological macromolecule, we create a minion as the sum of a set of 3D Gaussian distributions by separating different means of the Gaussian function in space similar to the 3D smiley face in Sec.~\ref{3DSmile}. The center of each Gaussian function is the position of atoms and we seek to reconstruct the electron density in these atoms. As Fig.~\ref{fig:minions_real} shows, the colorbar from yellow to red indicates the magnitude of Gaussian function at certain points which represents the density in real space. We also use the magnitude to set the size of each points in this space for better visual effect. We can clearly see the different body parts of this character. Actually, the whole space is full of points indicating the density but when the density gets very low, the points there are too small to be visible. Hence, most atoms center at the outline of the minion.

\begin{figure}[!h]
\centering
\begin{minipage}[t]{0.32\textwidth}
\centering
\includegraphics[width=5cm]{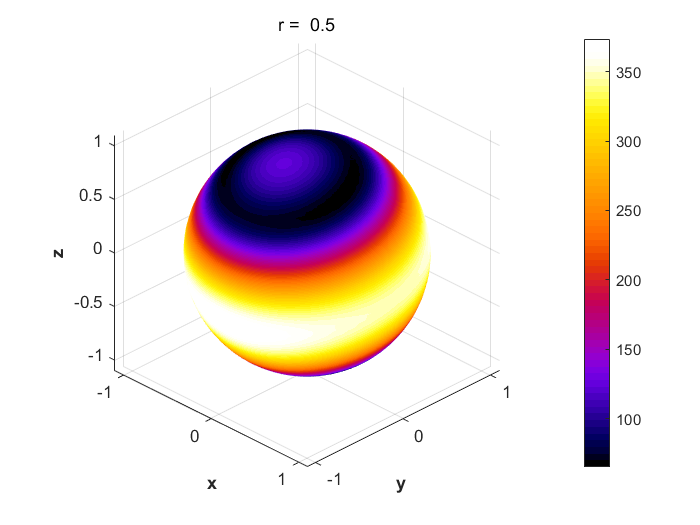}
\end{minipage}
\begin{minipage}[t]{0.32\textwidth}
\centering
\includegraphics[width=5cm]{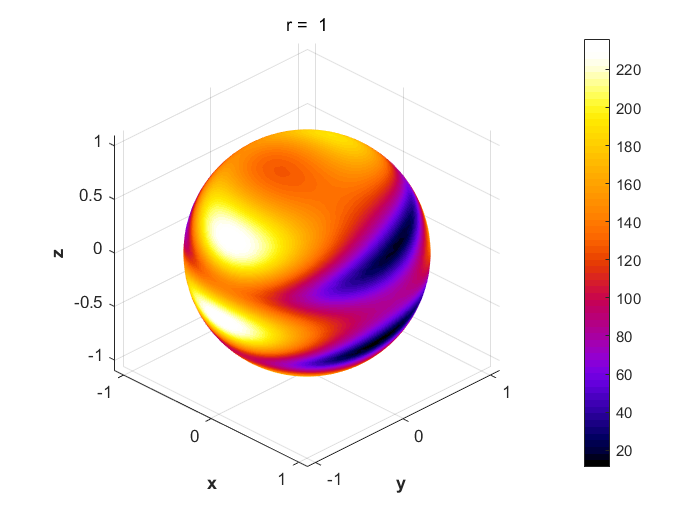}
\end{minipage}
\begin{minipage}[t]{0.32\textwidth}
\centering
\includegraphics[width=5cm]{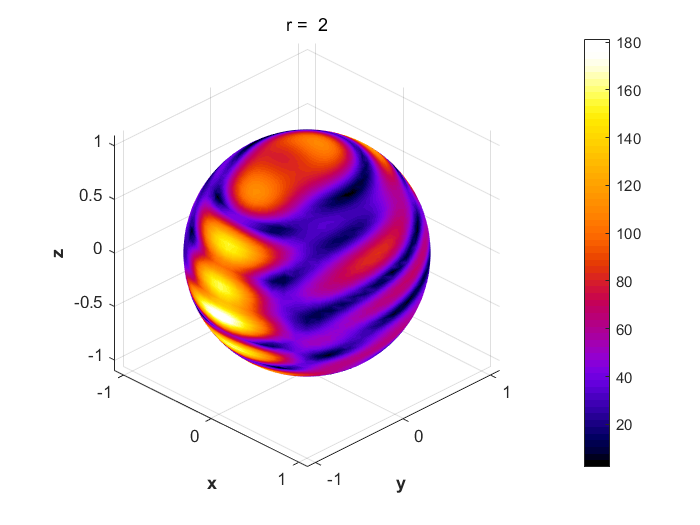}
\end{minipage}
\caption{Fourier transform of minion in spheres of radius 0.5, 1 and 2}
\label{Fig:Minions_fourier}
\end{figure}

\begin{figure}[!h]
    \centering
    \includegraphics[width=8.3cm]{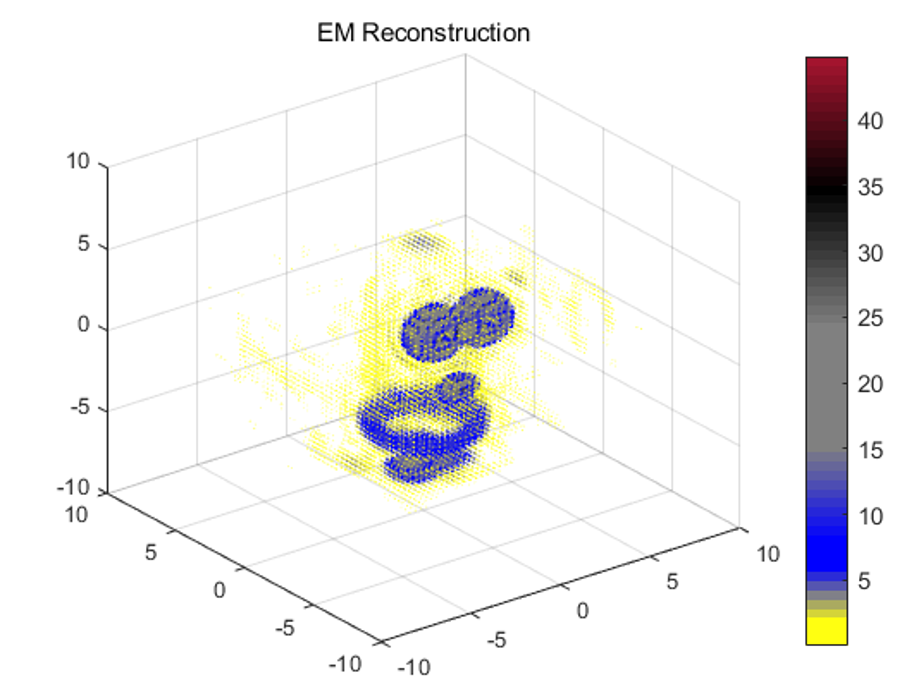}
    \caption{Reconstruction of Minion without exclusion of sampling points}
    \label{fig:minions_noremove_em}
\end{figure}

Figure \ref{Fig:Minions_fourier} shows what the minion is like in Fourier space. We pick spheres with radius 0.5, 1 and 2 to illustrate its Fourier transform magnitude. Similar to what we have experimented in 3D smiley face first, Fig.~\ref{fig:minions_noremove_em} shows the reconstruction results of the minion without cones to exclude sampling points in Fourier space. 
\FloatBarrier

\FloatBarrier

\begin{figure}[!h]
    \centering
    \includegraphics[width=13cm]{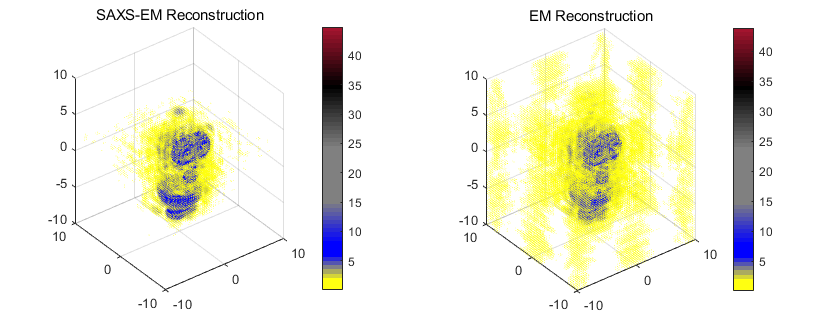}
    \caption{Comparison of Minion reconstruction by SAXS-EM fusion and EM-alone}
    \label{fig:minions_result1}
\end{figure}

And then cones are set in the sphere to exclude all the great circles whose normal direction are inside these cones in Fourier space. The center axis direction vector is defined in Sec. 5.3.2 and the great circles are generated uniformly in space. We still keep the radius dependent sampling number on each great circle here. The reconstruction result is shown in Fig.~\ref{fig:minions_result1}. 

\FloatBarrier

SAXS-EM fusion successfully restores the main body of the minion and especially the two hands which are the two clouds of black points on both sides of the body. Although we can see fuzzy blue eyes and feet by EM reconstruction alone, the main yellow body is separated in the whole space. It means the places where density should be extremely small are given comparatively higher density. 


\newpage
\section{Discussion, Conclusion, and Outlook}
\label{sec:discussion_and_conclusion}
In this paper, we presented a novel method of fusing information obtained from Small Angle X-ray Scattering (SAXS) and cryo-EM experiments. Despite the fact that SAXS data is of a lower resolution than EM data, it can be shown that information fusion greatly improves an EM reconstruction especially when the EM data is sparse or misses important regions of high information density. Numerical experiments with shapes in two and three-dimensions demonstrated the advantages of implementing a SAXS-EM fusion as opposed to an EM-only reconstruction. On the other hand, there also exists a theoretical lower bound for the minimum intensity that a given sample of EM data can accommodate and the SAXS information fusion would only add new information if the intensity is above this value for a given radius in Fourier space. Finally, we make the following set of observations that we hope would spur further research in the capabilities of SAXS-EM fusion and particularly, with applications to problems involving the elucidation of macromolecule structure from sparse EM data:
\begin{enumerate}
    \item Application of the SAXS-EM fusion to complex biological macromolecules. In this paper, we tested the SAXS-EM fusion with a model of a ``minion", with encouraging results. It reveals that the SAXS-EM fusion technique would be useful to obtain important information that EM experiments may otherwise miss by virtue of having the slice planes not pass through important structural details of biological macromolecules. Since the SAXS information is a spherical average, outstanding structural information would be recorded and can be used to enrich the EM data.
    \item The matrices involved in the solution of the quadratic equation tend to become ill-conditioned when the angles sampled are too close to each other. To allow the possibility of finely sampled EM data, the solution method to obtain the roots for the governing quadratic equation may need improvement so as to avoid the need to deal with ill-conditioned matrices.
    \item Allowing the number of sample points on a given great circle to vary linearly with the radius of the sphere leads to large improvements in the SAXS-EM reconstruction, and this technique can be utilized in further implementations of SAXS-EM fusion.
    \item The root selection procedure proposed in the paper was computationally efficient and also accurate. Nevertheless, it may also be possible to formulate more robust selection rules through integral quadratures (specifically, the stable and accurate integration over scattered data on a sphere) and allow for a more accurate identification of spurious roots. In fact, it may be possible to add further terms to ``correct" the roots obtained from the quadratic equation to obtain more accurate results.
\end{enumerate}

\section*{Conflicts of Interest}
The authors declare that there is no conflict of interest regarding the publication of this article.

\section*{Acknowledgement}
Research reported in this publication was supported by the National Institute of General Medical Sciences of the National Institutes of Health under award number R01GM113240.

\section*{Appendix A - The $N$-dimensional Gaussian}

The definition of the $N$-dimensional Gaussian is reviewed in the following. Shapes in two-dimensions and three-dimensions can be constructed as a superposition of two-dimensional and three-dimensional Gaussians respectively. The $N$-dimensional Gaussian function defined at $\bm{x}$, and with a mean of $\bm{\mu}$ and covariance matrix, $\Sigma$, is,

\begin{equation}\label{NGaussian}
f(\bm{x} ;\bm{\mu}, \Sigma)=\frac{1}{(2 \pi)^{\frac{N}{2}}|\operatorname{det} \Sigma|^{\frac{1}{2}}} \exp \left\{-\frac{1}{2}(\mathrm{x}-\boldsymbol{\mu})^{T} \Sigma^{-1}(\mathrm{x}-\boldsymbol{\mu})\right\}
\end{equation}

The Fourier transform of the Gaussian function is also a Gaussian function in Fourier space, and can be written as a function of $\bm{\omega}$ as,

\begin{equation}\label{NGaussian_Fourier}
\hat{f}(\bm{\omega} ;\bm{\mu}, \Sigma)= \exp\left(-\frac{\bm{\omega}^T\,\Sigma\,\bm{\omega}}{2} - i\bm{\mu}\right).
\end{equation}

Hence, it is possible to approximate a general shape $\phi(\bm{x})$ as a linear combination of $n$ Gaussians through,

\begin{equation}\label{NGaussian_Combine}
\phi(\bm{x})\approx\sum^{n}_{k = 1}A_k\,f(\bm{x};\bm{\mu}_k,\Sigma_k)
\end{equation}

\noindent where $A_k$ is the area of the $k^{\textrm{th}}$ Gaussian function, and used to control the weight of the Gaussian and thereby alter its contribution to the total sum. Since the Fourier transform is a linear operator, the Fourier transform of $\phi(\bm{x})$ would be,

\begin{equation}\label{NGaussian_Fourier_Combine}
\hat{\phi}(\bm{\omega})\approx\sum^{n}_{k = 1}A_k\,\hat{f}(\bm{\omega};\bm{\mu}_k,\Sigma_k).
\end{equation}

\bibliography{references}

\begin{thebibliography}{}

\bibitem[\protect\astroncite{Afsari et~al.}{2015}]{afsari_kim_chirikjian}
Afsari, B., Kim, {J.\,S.}, and Chirikjian, {G.\,S.} (2015).
\newblock {Cross-validation of data in SAXS and cryo-EM}.
\newblock In {\em {\textnormal{IEEE} International Conference on Bioinformatics
  and Biomedicine \textnormal{(BIBM)}}}, pages 1224--1230.

\bibitem[\protect\astroncite{Bhamre et~al.}{2016}]{bhamre2016denoising}
Bhamre, Tejal, Zhang, Teng, and Singer, Amit (2016).
\newblock Denoising and covariance estimation of single particle cryo-em
  images.
\newblock {\em Journal of structural biology}, 195(1):72--81.

\bibitem[\protect\astroncite{Biedenharn and Louck}{1981}]{biedenharn_louck}
Biedenharn, {L.\,C.}, and Louck, {J.\,D.} (1981).
\newblock {\em {Angular Momentum in Quantum Physics: Theory and Application}}.
\newblock Addison-Wesley, Boston, MA, USA.

\bibitem[\protect\astroncite{Blanchet and Svergun}{2013}]{blanchet2013small}
Blanchet, Clement~E, and Svergun, Dmitri~I (2013).
\newblock Small-angle x-ray scattering on biological macromolecules and
  nanocomposites in solution.
\newblock {\em Annual review of physical chemistry}, 64:37--54.

\bibitem[\protect\astroncite{Chirikjian and
  Kyatkin}{2016}]{chirikjian2016harmonic}
Chirikjian, Gregory~S, and Kyatkin, Alexander~B (2016).
\newblock {\em Harmonic Analysis for Engineers and Applied Scientists: Updated
  and Expanded Edition}.
\newblock Courier Dover Publications.

\bibitem[\protect\astroncite{Crowther
  et~al.}{1970}]{crowther1970reconstruction}
Crowther, Richard~Anthony, DeRosier, DJ, and Klug, Aaron (1970).
\newblock The reconstruction of a three-dimensional structure from projections
  and its application to electron microscopy.
\newblock {\em Proceedings of the Royal Society of London. A. Mathematical and
  Physical Sciences}, 317(1530):319--340.

\bibitem[\protect\astroncite{Dong et~al.}{2015a}]{dong_kim_chirikjian}
Dong, H., Kim, {J.\,S.}, and Chirikjian, {G.\,S.} (2015a).
\newblock {Computational analysis of SAXS data acquisition}.
\newblock {\em {J.\ Comput.\ Biol.}}, 22(9):787--805.

\bibitem[\protect\astroncite{Dong et~al.}{2015b}]{dong2015computational}
Dong, Hui, Kim, Jin~Seob, and Chirikjian, Gregory~S (2015b).
\newblock Computational analysis of saxs data acquisition.
\newblock {\em Journal of Computational Biology}, 22(9):787--805.

\bibitem[\protect\astroncite{Feigin et~al.}{1987}]{feigin1987structure}
Feigin, LA, Svergun, Dimitrij~I et~al. (1987).
\newblock {\em Structure analysis by small-angle X-ray and neutron scattering},
  Volume~1.
\newblock Springer.

\bibitem[\protect\astroncite{Frank}{2006}]{frank2006three}
Frank, Joachim (2006).
\newblock {\em Three-dimensional electron microscopy of macromolecular
  assemblies: visualization of biological molecules in their native state}.
\newblock Oxford University Press.

\bibitem[\protect\astroncite{Freeden et~al.}{1998}]{freeden_gervens_schreiner}
Freeden, Willi, Gervens, Theodor, and Schreiner, Michael (1998).
\newblock {\em {Constructive Approximation on the Sphere}}.
\newblock {Oxford University Press}, Oxford, UK.

\bibitem[\protect\astroncite{Kim et~al.}{2017}]{kim2017cross}
Kim, Jin~Seob, Afsari, Bijan, and Chirikjian, Gregory~S (2017).
\newblock Cross-validation of data compatibility between small angle x-ray
  scattering and cryo-electron microscopy.
\newblock {\em Journal of Computational Biology}, 24(1):13--30.

\bibitem[\protect\astroncite{Kunis and Potts}{2003}]{kunis_potts}
Kunis, Stefan, and Potts, Daniel (2003).
\newblock {Fast spherical Fourier algorithms}.
\newblock {\em {J.\ Comput.\ Appl.\ Math.}}, 161(1):75--98.

\bibitem[\protect\astroncite{Park and Chirikjian}{2014}]{park2014assembly}
Park, Wooram, and Chirikjian, Gregory~S (2014).
\newblock An assembly automation approach to alignment of noncircular
  projections in electron microscopy.
\newblock {\em IEEE Transactions on Automation Science and Engineering},
  11(3):668--679.

\bibitem[\protect\astroncite{Park et~al.}{2011}]{park2011stochastic}
Park, Wooram, Midgett, Charles~R, Madden, Dean~R, and Chirikjian, Gregory~S
  (2011).
\newblock A stochastic kinematic model of class averaging in single-particle
  electron microscopy.
\newblock {\em The International journal of robotics research}, 30(6):730--754.

\bibitem[\protect\astroncite{Penczek et~al.}{1992}]{penczek1992three}
Penczek, Pawel, Radermacher, Michael, and Frank, Joachim (1992).
\newblock Three-dimensional reconstruction of single particles embedded in ice.
\newblock {\em Ultramicroscopy}, 40(1):33--53.

\bibitem[\protect\astroncite{Penczek}{2002}]{penczek2002three}
Penczek, Pawel~A (2002).
\newblock Three-dimensional spectral signal-to-noise ratio for a class of
  reconstruction algorithms.
\newblock {\em Journal of structural biology}, 138(1-2):34--46.

\bibitem[\protect\astroncite{Penczek et~al.}{1996}]{penczek1996common}
Penczek, Pawel~A, Zhu, Jun, and Frank, Joachim (1996).
\newblock A common-lines based method for determining orientations for n> 3
  particle projections simultaneously.
\newblock {\em Ultramicroscopy}, 63(3-4):205--218.

\bibitem[\protect\astroncite{Potts et~al.}{2001}]{potts_steidl_tasche}
Potts, Daniel, Steidl, Gabriele, and Tasche, Manfred (2001).
\newblock {Fast Fourier transforms for nonequispaced data: a tutorial}.
\newblock In Benedetto, John~J., and Ferreira, Paulo~{J.\,S.\,G.}, editors,
  {\em Modern Sampling Theory}, Applied and Numerical Harmonic Analysis, pages
  247--270. Birkh\"auser Boston, MA, USA.

\bibitem[\protect\astroncite{Schatz and Van~Heel}{1990}]{schatz1990invariant}
Schatz, Michael, and Van~Heel, Marin (1990).
\newblock Invariant classification of molecular views in electron micrographs.
\newblock {\em Ultramicroscopy}, 32(3):255--264.

\bibitem[\protect\astroncite{Scheres}{2012}]{scheres2012relion}
Scheres, Sjors~HW (2012).
\newblock Relion: implementation of a bayesian approach to cryo-em structure
  determination.
\newblock {\em Journal of structural biology}, 180(3):519--530.

\bibitem[\protect\astroncite{Scheres et~al.}{2005}]{scheres2005maximum}
Scheres, Sjors~HW, Valle, Mikel, Nu{\~n}ez, Rafael, Sorzano, Carlos~OS,
  Marabini, Roberto, Herman, Gabor~T, and Carazo, Jose-Maria (2005).
\newblock Maximum-likelihood multi-reference refinement for electron microscopy
  images.
\newblock {\em Journal of molecular biology}, 348(1):139--149.

\bibitem[\protect\astroncite{Sherman and Morrison}{1950}]{Sherman1950}
Sherman, Jack, and Morrison, Winifred~J. (1950).
\newblock Adjustment of an inverse matrix corresponding to a change in one
  element of a given matrix.
\newblock {\em The Annals of Mathematical Statistics}, 21(1):124--127.

\bibitem[\protect\astroncite{Shkolnisky and
  Singer}{2012}]{shkolnisky2012viewing}
Shkolnisky, Yoel, and Singer, Amit (2012).
\newblock Viewing direction estimation in cryo-em using synchronization.
\newblock {\em SIAM journal on imaging sciences}, 5(3):1088--1110.

\bibitem[\protect\astroncite{Sigworth}{1998}]{sigworth1998maximum}
Sigworth, Fred~J (1998).
\newblock A maximum-likelihood approach to single-particle image refinement.
\newblock {\em Journal of structural biology}, 122(3):328--339.

\bibitem[\protect\astroncite{Singer et~al.}{2011}]{singer2011viewing}
Singer, Amit, Zhao, Zhizhen, Shkolnisky, Yoel, and Hadani, Ronny (2011).
\newblock Viewing angle classification of cryo-electron microscopy images using
  eigenvectors.
\newblock {\em SIAM Journal on Imaging Sciences}, 4(2):723--759.

\bibitem[\protect\astroncite{Stuhrmann}{1970}]{stuhrmann1970interpretation}
Stuhrmann, HEINRICH~B (1970).
\newblock Interpretation of small-angle scattering functions of dilute
  solutions and gases. a representation of the structures related to a
  one-particle scattering function.
\newblock {\em Acta Crystallographica Section A: Crystal Physics, Diffraction,
  Theoretical and General Crystallography}, 26(3):297--306.

\bibitem[\protect\astroncite{Svergun and Koch}{2003}]{svergun2003small}
Svergun, Dmitri~I, and Koch, Michel~HJ (2003).
\newblock Small-angle scattering studies of biological macromolecules in
  solution.
\newblock {\em Reports on Progress in Physics}, 66(10):1735.

\bibitem[\protect\astroncite{Van~Heel and Frank}{1981}]{van1981use}
Van~Heel, Marin, and Frank, Joachim (1981).
\newblock Use of multivariates statistics in analysing the images of biological
  macromolecules.
\newblock {\em Ultramicroscopy}, 6(1):187--194.

\bibitem[\protect\astroncite{Wang and Sigworth}{2006}]{wang2006cryo}
Wang, Liguo, and Sigworth, Fred~J (2006).
\newblock Cryo-em and single particles.
\newblock {\em Physiology}, 21(1):13--18.

\end{thebibliography}
\bibliographystyle{my_apa}

\end{document}